\documentclass[prb,superscriptaddress, preprintnumbers, twocolumn, letterpaper]{revtex4}

\usepackage{times}
\usepackage{amsmath}
\usepackage{amsthm}
\usepackage{amssymb}
\usepackage{amsbsy}
\usepackage{amsfonts}
\usepackage{graphicx}
\usepackage{tikz}

\def\pmat#1{{\begin{pmatrix}#1\end{pmatrix}}}

\begin{document}



\title{The Partially-Split Hall Bar: Tunneling in the Bosonic Integer Quantum Hall Effect}

\author{Michael Mulligan}
\affiliation{Microsoft Research, Station Q, Elings Hall,
University of California, Santa Barbara, California 93106-6105, USA}
\author{Matthew P. A. Fisher}
\affiliation{Department of Physics, University of California, Santa Barbara,
California 93106, USA}

\begin{abstract}
We study point-contact tunneling in the integer quantum Hall state of bosons.
This symmetry-protected topological state has electrical Hall conductivity equal to $2 e^2/h$ and vanishing thermal Hall conductivity.
In contrast to the integer quantum Hall state of fermions, a point contact can have a dramatic effect on the low energy physics.
In the absence of disorder, a point contact generically leads to a {\it partially}-split Hall bar geometry.
We describe the resulting intermediate fixed point via the two-terminal electrical (Hall) conductance of the edge modes.
Disorder along the edge, however, both restores the universality of the two-terminal conductance and helps preserve the integrity of the Hall bar within the relevant parameter regime.
\end{abstract}

\maketitle

\section{Introduction}

The low-energy excitations of a quantum Hall droplet live along the edges of the sample.
When two opposing edges of the same droplet are well separated, the bulk mobility gap prevents tunneling interactions between these low-energy excitations.
However, if a constriction is introduced such that the two opposing sides meet near a point, the amplitude for inter-edge tunneling can be appreciable.
At finite temperature, the resulting quasiparticle tunneling degrades the Hall current as it allows backscattering between opposing edges at the point contact.

As an example, consider the Laughlin states at filling fraction $\nu  = 1/m$.
These states have (fractionally) quantized Hall conductance $G_0(T) = \nu e^2/h$ in the absence of a constriction.
Tunneling at any such constriction is generically dominated by the transfer of (fractionally-)charged $\nu e$ quasiparticles and leads to the reduction of the Hall current by the amount, $G_a(T) - \nu e^2/h = - a T^{2 \nu - 2}$ for some positive constant $a$.\cite{kanefishertunnel}
While this perturbative result is necessarily only valid for temperatures, $T^{2 - 2 \nu} \gg {ah \over \nu e^2} $, it shows the marked difference between the integral and fractional quantum Hall regimes when extrapolated to zero temperature.
For the integral case, the conductance is merely reduced by a finite, constant value that is independent of temperature.
In the fractional case, $2 - 2 \nu > 0$, so the perturbative calculation indicates that the effect of backscattering at the point contact on the Hall conductance is to reduce its value as the temperature is lowered, eventually leading to a vanishing Hall conductance.
This means that the Hall droplet has effectively split into two pieces.
The perturbative picture is supported by an exact solution at certain filling fractions: the theoretical model that describes the edge mode conductance is integrable at $\nu = 1/3$ and at other Laughlin states after fine-tuning to a single relevant interaction.\cite{FLSshort, FLSlong}

\begin{figure}[t]
\begin{tikzpicture}
\node at (-9,0) {{\bf 1A.}};
\node at (-7,0) {{\scriptsize Vacuum}};
\node at (-7, -3) {{\scriptsize Vacuum}};
\node at (-8, -1.5) {{\scriptsize Hall Fluid}};
\node at (-6, -1.5) {{\scriptsize Hall Fluid}};
\node at (-2, 0) {{\scriptsize Vacuum}};
\node at (-2,-3) {{\scriptsize Vacuum}};
\node at (-3, -1.5) {{\scriptsize Hall Fluid}};
\node at (-1, -1.5) {{\scriptsize Hall Fluid}};
\node at (-6.58, -4.25) {{\scriptsize Vacuum}};
\node at (-6.58, -6.25) {{\scriptsize Vacuum}};
\node at (-5, -5.25) {{\scriptsize Hall Fluid}};
\node at (-8, -5.25) {{\scriptsize Hall Fluid}};
\draw [->] (-8,0) -- (-7.5,-.5);
\draw (-7.5,-.5) -- (-7,-1);
\draw [->] (-7,-1) -- (-6.5, -.5);
\draw (-6.5, -.5) -- (-6,0);
\draw (-8.25, 0) -- (-7.5, -.75);
\draw [<-] (-7.5, -.75) -- (-7, -1.25);
\draw (-7,-1.25) -- (-6.5, -.75);
\draw [<-] (-6.5, -.75) -- (-5.75, 0);
\draw [dashed, thick, blue] (-7.25, -1.25) -- (-7.25, -1.5);
\draw [<-] [dashed,thick, blue] (-7.25 ,-1.5) -- (-7.25, -1.75);
\draw [->] [dashed,thick, blue] (-6.75, -1.25) -- (-6.75, -1.5);
\draw [dashed,thick, blue] (-6.75, -1.75) -- (-6.75, -1.5);
\draw [->] (-8.25, -3) -- (-7.5, -2.25);
\draw (-7.5, -2.25) -- (-7, -1.75);
\draw [->] (-7,-1.75) -- (-6.5, -2.25);
\draw (-6.5, -2.25) -- (-5.75, -3);
\draw (-8, -3) -- (-7.5, -2.5);
\draw [<-] (-7.5, -2.5) -- (-7,-2);
\draw  (-7,-2) -- (-6.5, -2.5);
\draw [<-] (-6.5, -2.5) -- (-6,-3);
\node at (-4, 0) {{\bf 1B.}};
\draw [->] (-3,0) -- (-2.5,-.5);
\draw (-2.5,-.5) -- (-2,-1);
\draw [->] (-2,-1) -- (-1.5, -.5);
\draw (-1.5, -.5) -- (-1,0);
\draw (-3.25, 0) -- (-2.5, -.75);
\draw [<-] (-2.5, -.75) -- (-2, -1.25);
\draw  (-2,-1.25) -- (-1.5, -.75);
\draw [<-] (-1.5, -.75) -- (-0.75, 0);
\draw [->] (-3.25, -3) -- (-2.5, -2.25);
\draw (-2.5, -2.25) -- (-2, -1.75);
\draw [->] (-2,-1.75) -- (-1.5, -2.25);
\draw (-1.5, -2.25) -- (-0.75, -3);
\draw (-3, -3) -- (-2.5, -2.5);
\draw [<-] (-2.5, -2.5) -- (-2,-2);
\draw  (-2,-2) -- (-1.5, -2.5);
\draw [<-] (-1.5, -2.5) -- (-1,-3);
\draw [fill=red, ultra thick,green] (-2.25,-.75) rectangle (-1.75,-2.25);
\node at (-9,-4) {{\bf 1C.}};
\draw [->] (-8,-4) -- (-7.5,-4.5);
\draw (-7.5,-4.5) -- (-6.75,-5.25);
\draw [->] (-6.25,-5.25) -- (-5.5, -4.5);
\draw (-5.5, -4.5) -- (-5,-4);
\draw (-8.25, -4) -- (-7.5, -4.75);
\draw [<-] (-7.5, -4.75) -- (-7, -5.25);
\draw (-6,-5.25) -- (-5.5, -4.75);
\draw [<-] (-5.5, -4.75) -- (-4.75, -4);
\draw [->] (-8.25, -6.5) -- (-7.5, -5.75);
\draw (-7.5, -5.75) -- (-7, -5.25);
\draw [->] (-6,-5.25) -- (-5.5, -5.75);
\draw (-5.5, -5.75) -- (-4.75, -6.5);
\draw (-8, -6.5) -- (-7.5, -6);
\draw [<-] (-7.5, -6) -- (-6.75,-5.25);
\draw  (-6.25,-5.25) -- (-5.5, -6);
\draw [<-] (-5.5, -6) -- (-5,-6.5);
\draw [->] [dashed, thick, red] (-6.75, -5) -- (-6.5, -5);
\draw [dashed, thick, red] (-6.5, -5) -- (-6.25, -5);
\draw [dashed, thick,red] (-6.75, -5.5) -- (-6.5, -5.5);
\draw [<-] [dashed, thick, red] (-6.5, -5.5) -- (-6.25, -5.5);
\end{tikzpicture}
\caption{Qualitative geometries of the Hall bar at the various fixed points.
The fully connected geometry in Fig. 1A describes a single-component Hall bar.
The partially-split geometry in Fig. 1B represents the intermediate fixed point present in certain parameter regimes in the bosonic IQH state, spin Hall insulator, and the bosonic $E_8$ state.
The fully disconnected geometry in Fig. 1C describes a Hall bar consisting of two disconnected components coupled via a point-contact interaction.}
\label{figone}
\end{figure}
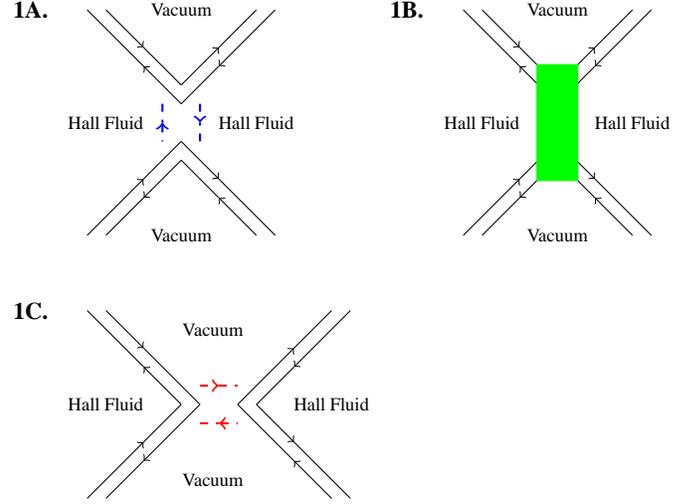

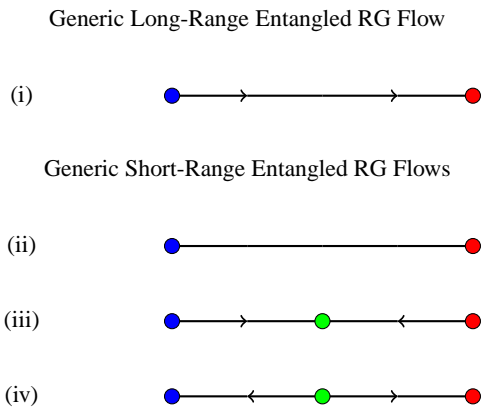
\begin{figure}[t]
\begin{tikzpicture}
\node at (-1,0) {Generic Long-Range Entangled RG Flow};
\node at (-4,-1) {(i)};
\draw [->] [thick] (-2,-1) -- (-1,-1);
\draw [fill = blue] (-2,-1) circle [radius=0.1];
\draw [thick] (-1,-1) -- (0,-1);
\draw [->] [thick] (0,-1) -- (1,-1);
\draw [thick] (1,-1) -- (2,-1);
\draw [fill = red] (2,-1) circle [radius = .1];
\node at (-1,-2) {Generic Short-Range Entangled RG Flows};
\node at (-4,-3) {(ii)};
\draw  [thick] (-2,-3) -- (-1,-3);
\draw [fill = blue] (-2,-3) circle [radius=0.1];
\draw [thick] (-1,-3) -- (0,-3);
\draw [thick] (0,-3) -- (1,-3);
\draw [thick] (1,-3) -- (2,-3);
\draw [fill = red] (2,-3) circle [radius = .1];
\node at (-4,-4) {(iii)};
\draw  [thick] [->] (-2,-4) -- (-1,-4);
\draw [fill = blue] (-2,-4) circle [radius=0.1];
\draw [thick] (-1,-4) -- (0,-4);
\draw [thick] (0,-4) -- (1,-4);
\draw [fill=green] (0,-4) circle [radius=.1];
\draw [thick] [<-] (1,-4) -- (2,-4);
\draw [fill = red] (2,-4) circle [radius = .1];
\node at (-4,-5) {(iv)};
\draw  [thick] (-2,-5) -- (-1,-5);
\draw [fill = blue] (-2,-5) circle [radius=0.1];
\draw [thick] [<-] (-1,-5) -- (0,-5);
\draw [thick][->] (0,-5) -- (1,-5);
\draw [fill=green] (0,-5) circle [radius=.1];
\draw [thick] (1,-5) -- (2,-5);
\draw [fill = red] (2,-5) circle [radius = .1];
\end{tikzpicture}
\caption{Qualitative boundary renormalization group flow diagrams of point-contact perturbed Hall fluids.
The (blue) left-most dot represents the fixed point describing a fully connected Hall bar geometry while the right-most (red) dot represents the fixed point that describes a Hall bar geometry that has fully split into two pieces.
A (green) dot in between represents an intermediate fixed point.
A fixed point is IR stable with respect to a particular (irrelevant) operator if the arrow on the line connected to it is directed towards the  representative dot, while a line with no arrow indicates a fixed line.
Fractional quantum Hall states realize (i).
Integer quantum Hall states of fermions realize the fixed line drawn in (ii).
The bosonic integer quantum Hall state -- the state that is the subject of this paper -- realizes (ii) and (iii) in the dirty and clean limits, respectively.
The bosonic $E_8$ state realizes (iv).
The spin Hall insulator realizes (ii) - (iv) in various edge interaction parameter regimes.}
\label{figtwo}
\end{figure}

The above results can be understood purely from the perspective of the low-energy edge modes.
Consider the situation where the Hall fluid is put on the infinite strip.
For the $\nu = 1/m$ Laughlin states, the counter-propagating edge modes living on each side of the strip together form a non-chiral Luttinger liquid.
The point contact provides a perturbation at a single spatial point that scatters a left moving mode into a right moving mode and vice versa.
The (boundary) scaling dimension of this tunneling operator $\Delta = \nu$ and is relevant, in the renormalization group (RG) sense, if $\nu < 1$.
If relevant, the perturbation drives the theory to a new infrared (IR) fixed point which, in the Laughlin case, corresponds to a fully disconnected geometry where the Hall bar has split in two.
When $\nu = 1$, the perturbation is exactly marginal and describes a line of fixed points parameterized by the coefficient of this perturbing operator.
In Fig. \ref{figone}A and Fig. \ref{figone}C, we draw the geometries corresponding to the fully connected and fully disconnected fixed points present in point-contact perturbed Hall states.
The analysis of the Laughlin states immediately generalizes to other abelian fractions with the basic conclusion being that an abelian fluid perturbed by a point contact will generically flow to a new IR fixed point if the abelian state has quasiparticles with fractional braid statistics.

From the edge point of view, the point contact is an impurity that leads at low energies to a change in the boundary conditions for the edge modes at the location of the impurity, if the tunneling operator is relevant.\cite{FLSshort, wongafflecktunnel}
In the simplest of cases, an example of which is provided by the state considered in this paper, the boundary conditions describe how left moving edge modes a reflected into right moving edge modes and vice versa.
Thus, a boundary RG flow is initiated by the presence of the point contact.
(A boundary RG flow is one where the edge mode dispersion remains gapless along the RG trajectory, while there is a change in the conformally-invariant boundary conditions between the ultraviolet (UV) and IR fixed points.\cite{Cardybcfusionverlinde})
This flow is characterized by the change in the Affleck-Ludwig boundary entropy \cite{AffleckLudwigentropy}; the boundary entropy is a scalar quantity that functions much like the central charge \cite{Zamolodchikovctheorem} in 1+1d conformal field theory (CFT) as it monotonically decreases along boundary RG trajectories.\cite{FriedanKonechny}

Using the celebrated bulk-boundary correspondence of Chern-Simons theory, Fendley, Fisher, and Nayak made a beautiful observation relating the behavior of the edge and bulk theories upon perturbation of a Hall state by a point contact.\cite{FFNholographic}
They identified the change in the boundary entropy with the change in the bulk thermodynamic entropy of the Hall state.
The change in the latter quantity is negative because a splitting of the Hall bar implies a decrease in the amount of uncertainty regarding the state of the fluid.
The change in both quantities is equal to $- \log({\cal D})$, where the total quantum dimension of a topological state, ${\cal D} = \sqrt{\sum_i d_i^2} = 1/S_{00} \geq 1$, where $d_i$ are the quantum dimensions of the individual quasiparticles of a topological state and $S_{00}$ is the $00$-th entry of the modular S-matrix.
This identification unites the bulk and boundary viewpoints on the behavior of the Hall fluid upon perturbation by a point contact.

In this paper, a long-range entangled (topological) state of matter is defined to be a gapped state with non-zero topological entanglement entropy,\cite{hammakitaev, KitaevPreskillentropy, LevinWenentropy} $- \log({\cal D})$, and so ${\cal D} >1$; a short-range entangled (topological) state is then a gapped state with vanishing topological entanglement entropy, ${\cal D} = 1$.
Thus, in the absence of symmetry considerations, a point contact has a very different effect on Hall states that are long-range entangled versus those that are only short-range entangled: long-range entangled states generally split in two under RG flow while short-range entangled states do not.
A RG trajectory between the fully connected and fully disconnected limits is only possible when ${\cal D} >1$.
Otherwise, there must either be a fixed line connecting the fully connected and fully disconnected fixed points or there must be an intermediate fixed point that prevents a direct flow between these two limits.
To emphasize this point, we sketch in Fig. \ref{figtwo} the qualitative RG flows for a few illustrative quantum Hall liquids.

In 2+1d, short-range entangled states of fermions are built from layers of $\nu = 1$ Laughlin states which are characterized by their thermal Hall conductance \cite{kanefisherthermal} (in the absence of any symmetry), a quantity that measures the chiral central charge of the state.
For an abelian state, the chiral central charge is simply the difference in the number of left and right moving edge modes.
As we have reviewed, a point contact in this system is merely an exactly marginal perturbation and does not lead to a splitting of the Hall droplet as the temperature is lowered.
In the absence of symmetry, short-range entangled states of bosons are built from layers of the $E_8$ state which have a minimum of eight chiral edge modes.\cite{Kitaevanyonsexact, Kitaev11, LuVishwanathcsclass, PlamadealaE8}
In contrast to the $\nu =1$ Laughlin state, the lowest dimension point-contact tunneling operator for the $E_8$ state has (boundary) scaling dimension $\Delta = 2$ and is strictly irrelevant.
Again, the state does not split in two as the temperature is lowered.

Symmetry-protected topological states are short-range entangled states that are stabilized by a particular global symmetry.\cite{FuKaneinversion, QHZtop, xieliuwen, Chen11b}
If the symmetry is broken, the state is adiabatically connected to the trivial vacuum state without a closing of the bulk gap.
It is interesting to ask whether symmetry-protected states can display novel behavior that is not shared by short-range entangled or long-range entangled states obeying no symmetry requirements.

This question has been studied in the context of the time-reversal invariant spin Hall insulator.\cite{houkimchamon, teokanepointcontact}
These authors mapped the Lagrangian describing the edge modes of the spin Hall insulator to the Lagrangian of two decoupled Luttinger liquids of spin and charge bosons.
The Luttinger parameters, $g_c$ and $g_s$, cannot take arbitrary values; rather, they are constrained to lie upon the line $g_c \times g_s = 1$.
As we would expect, there is no RG flow between the fully connected and disconnected fixed points.
Away from the marginal point in parameter space at $g_c = g_s = 1$, there is always an intermediate fixed point that is IR stable or unstable, depending upon where on the line, $g_c \times g_s = 1$, the electron-electron interactions lie.

In this paper, we study the simplest bosonic symmetry-protected topological state that respects a $U(1)$ charge-conservation symmetry.\cite{LevinSenthilbosons, YeWen}
The electrical Hall conductivity of this bosonic state takes the minimum value $\sigma_{xy} = 2 e^2/h$ consistent with the fact that it is short-range entangled and can border the trivial vacuum.
Within this note, we refer to this state as the bosonic integer quantum Hall (IQH) state.
(A second candidate for this name is the previously mentioned $E_8$ bosonic state that has eight chiral edge modes.)
Interestingly and in contrast to the spin Hall insulator, we find that the point contact (almost) {\it always} generates an RG trajectory to a new fixed point that describes a Hall droplet that has {\it partially} split in two.
The RG diagram for this flow is shown in Fig. \ref{figtwo} (iii).
Here, we are assuming that the edge is clean and that we are working away from the single point in parameter space where the point-contact perturbation is marginal.
We characterize the resulting IR stable fixed point in terms of observable current-current correlation functions.
Surprisingly, we also find that the flow to this fixed point is integrable in the same sense that the RG flow of the point-contact perturbed fractional quantum Hall state is integrable.

Disorder along the edge modifies this picture.\cite{KFPrandom, KFrandom}
In the domain of attraction of the strong disorder fixed point, the leading point-contact tunneling term turns out to be {\it marginal}.
And so the point-contact perturbed Hall bar is described by the fixed line drawn in Fig. \ref{figtwo} (ii).
Disorder has the benefit of restoring the universality of the two-terminal electrical Hall conductance of the edge modes to its value inferred from the bulk topological order of the state.

The remainder of this note is organized as follows.
In Section \ref{preliminaries}, we introduce and define the bosonic IQH state.
In Section \ref{equilibration}, we describe how edge equilibration drives the system towards a strong-disorder fixed point where the two-terminal electrical Hall conductance equals $2 e^2/h$.
In Section \ref{pointcontact}, we add the point contact and describe the resulting fixed point for a clean edge.
In Section \ref{conductance}, we calculate various current-current correlation functions in order to characterize the different fixed points.
In Section \ref{summary}, we summarize and conclude.

\section{The Bosonic IQHE}
\label{preliminaries}

In the present section, we define the bulk and boundary actions that describe the bosonic IQH state in order to set the stage for the description of edge equilibration and point-contact tunneling in the following sections.

\subsection{Bulk Action}

By the bosonic IQH state, we mean the symmetry-protected topological state of bosons stabilized by $U(1)$ charge-conservation symmetry.\cite{LevinSenthilbosons, YeWen}
This state has vanishing thermal Hall conductivity, an electrical Hall conductivity $\sigma_{xy} = 2 e^2/h$, and the ability to border the topologically trivial vacuum.
This latter constraint means that the state can exist purely as a 2+1d topological phase and need not live on the boundary of a 3+1d spacetime.

The bosonic IQH state can be described using abelian Chern-Simons theory and the so-called K-matrix formalism.\cite{wenreview}
The bulk action,
\begin{align}
\label{edgelagnonchiral}
S_{{\rm bulk}} = \int d^2x dt \Big[ {\epsilon_{\mu \nu\rho} \over 4 \pi} K_{IJ} a_I^\mu \partial^\nu a_J^\rho - {\epsilon_{\mu \nu \rho} \over 2 \pi} t_I A^\mu \partial^\nu a_I^\rho \Big],
\end{align}
where the K-matrix, $K_{I J} = (\sigma_x)_{IJ} =  \pmat{0 & 1 \cr 1 & 0}$, and the charge-vector $t_I = (1,1)$.
Here, we are assuming the state is formed from bosonic constituents of charge $e$ that we set to unity unless otherwise specified.
The gauge fields $a_I^\mu$ describe the number currents of the bosons on each ``layer" indexed by $I=1,2$ and $A^\mu$ is a background electromagnetic field.
The spacetime index $\mu = 0,1,2 = t, x, y$.
Our convention for the totally anti-symmetric tensor $\epsilon_{\mu \nu \rho}$ is to choose $\epsilon_{0 1 2} = 1$.

The thermal Hall conductivity is equal to the signature of $K_{IJ}$ in units of $\pi^2 k_B^2 T/3h$.
In units of $e^2/h$, the electrical Hall conductivity $\sigma_{xy} = t_I (K^{-1})^{IJ} t_J = 2$.
Quasiparticle excitations are labelled by an integer vector $n_I = (n_1, n_2)$; their charge is given by $t_I (K^{-1})^{IJ} n_J = n_1 + n_2$ and their mutual statistics by $\exp(2 \pi i n'_I (K^{-1})^{IJ} n_J) = \exp(2 \pi i (n'_1 n_2 + n'_2 n_1))=1$.
The self-statistics of a quasiparticle $n_I$ is obtained by halving the expression for the mutual statistical angle and replacing $n'_I = n_I$,  thereby giving unity for any exchange.
Thus, all quasiparticles are bosonic and have integral charge.

\subsection{Edge Action}

The bulk Chern-Simons theory requires the presence of gapless edge modes living along the boundary of any space upon which the theory is defined.\cite{Elitzur89, Wen92edgereview}
We consider the bosonic IQH state to live on a strip of width $W$ in the y-direction and length $2L \rightarrow \infty$ in the x-direction.
Thus, there are two disconnected edges that we refer to as the top and bottom of the Hall bar located at, say, $y = W$ and $y= 0$, respectively.
For every bulk gauge field $a^\mu_I$ and for each boundary component, we introduce the bosonic edge modes $\phi_I$ and $\tilde{\phi}_I$.
Our convention is to take the $\phi_I$ fields to live on the top of the Hall bar and the $\tilde{\phi}_I$ fields to live along the bottom of the bar.
The two sets of edge modes $\phi_{I}$ and $\tilde{\phi}_{J}$ are used to describe the local tunneling physics, but not the global dynamics of the entire Hall droplet. 

The action for these edge modes,
\begin{align}
\label{startaction}
S_{{\rm edge}} & = {1 \over 4\pi} \int dt dx \Big[K_{I J} \partial_t \phi_I \partial_x \phi_J - V_{IJ} \partial_x \phi_I \partial_x \phi_J \cr
& +  \tilde{K}_{I J} \partial_t \tilde{\phi}_I \partial_x \tilde{\phi}_J - \tilde{V}_{IJ} \partial_x \tilde{\phi}_I \partial_x \tilde{\phi}_J \cr
& +   2 \epsilon^{\mu \nu} \Big( t_I A_\mu(W) \partial_\nu \phi_I + \tilde{t}_I A_\mu(0) \partial_\nu \tilde{\phi}_I \Big) \Big]
\end{align}
where $K_{I J} = - \tilde{K}_{I J} = \pmat{0 & 1 \cr 1 & 0}$, $t_I = \tilde{t}_I = \pmat{1 & 1}$, and $V_{I J}, \tilde{V}_{I J}$ are symmetric, positive-definite matrices that parameterize the density-density and current-current or forward scattering interactions along a given edge.
Clearly, the kinetic structure (and, therefore, the operator algebra) of the theory defined by the $K_{IJ}, \tilde{K}_{IJ}$ matrices is inherited from the bulk topological order, while the $V_{IJ}, \tilde{V}_{IJ}$ interactions are non-universal from the perspective of the bulk physics as they depend upon edge properties.
The edge modes are periodically identified,
\begin{align}
\label{periodicity}
\phi_I \sim \phi_I + 2 \pi a_I, \quad \tilde{\phi}_I \sim \tilde{\phi}_I + 2 \pi b_I,\ a_I, b_I \in \mathbb{Z}.
\end{align}
The background electromagnetic gauge field $A_\mu$ propagates throughout the bulk and we denote its restriction to a given edge by $A_\mu(y)$ for $y = 0, W$, while suppressing its $t$ and $x$ coordinates.
$A_\mu$ couples to the edge current densities, $j_I^\mu = {\epsilon^{\mu \nu} \over 2\pi} \partial_\nu \phi_I$ and $\tilde{j}_I^\mu = {\epsilon^{\mu \nu} \over 2\pi} \partial_\nu \tilde{\phi}_I$.

Because the signature of $K_{IJ}$ vanishes, the general expectation is that a symmetry is required to stabilize the modes on a given edge from the generation of a mass gap.
While there are exceptions to this rule,\cite{Haldanestability, Levin13, wangwen12, barkeshlijianqiboundary} the edge of the bosonic IQH state is not one of them.
Charge conservation protects the edge modes from the possible generation of a gap via backscattering.
Indeed, a general backscattering operator on the top edge has the form, $\cos(n_1 \phi_1 + n_2 \phi_2)$.
This operator carries total charge $t_I (K^{-1})^{IJ} n_J = n_1+ n_2$ and so we require $n_1 = - n_2$ for neutrality.
However, a mass-generating term must have equal left $\Delta_L$ and right $\Delta_R$ scaling dimensions; that is, such a term must have vanishing spin.
The field redefinitions in Eqns. (\ref{leftrightchirals}) and the action Eqn. (\ref{actionfull}) show that $\cos(n(\phi_1 - \phi_2))$ has spin equal to $\Delta_R - \Delta_L = - n^2$.
So we conclude that the edge of the bosonic IQH state is stable as long as charge conservation is maintained.
Phrased in terms of the null vector criterion,\cite{Haldanestability} the edge is stable because there does not exist a nontrivial neutral null vector for the bosonic IQH state.

In the remainder of the paper, we make the simplifying assumption that $V_{IJ} = \tilde{V}_{IJ}$.
In general, there is no relation between $V_{I J}$ and $\tilde{V}_{I J}$ when the two edges are disconnected.
However, this assumption is reasonable in an actual sample that has one connected boundary.
As will be shown in the next section, this assumption with $V_{IJ} = \delta_{IJ}$ is an RG attractor in the presence of relevant disorder scatterers.
Later on, we comment upon the fate of our proposed phase diagram when $V_{IJ} \neq \tilde{V}_{IJ}$.
However, we leave to future work the thorough investigation of the dynamics of the point-contact tunneling when the simplification $V_{IJ} = \tilde{V}_{IJ}$ is relaxed.

To study the effects of the density-density and current-current interactions parameterized by $V_{IJ}$ on the physics of a point contact, it is convenient to perform a field redefinition and rewrite $S_{{\rm edge}}$ in terms of left and right moving chiral fields.
We assume $V_{11}$ and $V_{22}$ are positive.
We then define:
\begin{align}
\label{leftrightchirals}
\phi_1 & = {1 \over \sqrt{2 g}}(X_R + X_L), \quad \phi_2 = {\sqrt{g} \over \sqrt{2}} (X_R - X_L),\cr
\tilde{\phi}_1 & = {1 \over \sqrt{2 g}}(\tilde{X}_R + \tilde{X}_L), \quad \tilde{\phi}_2 = {\sqrt{g} \over \sqrt{2}}(- \tilde{X}_R + \tilde{X}_L),
\end{align}
where $g = \sqrt{V_{11}/V_{22}}$  and $X^R, \tilde{X}^R$ ($X^L, \tilde{X}^L$) are functions of $x-t$ ($x+t$). 
In terms of the left and right moving fields, $S_{{\rm edge}}$ becomes,
\begin{align}
\label{actionfull}
S_{{\rm edge}} & = {1 \over 4 \pi} \int dt dx \Big[\partial_x X_R (\partial_t - v_R \partial_x) X_R \cr
& + \partial_x X_L (- \partial_t - v_L \partial_x) X_L  + \partial_x \tilde{X}_R (\partial_t - v_R \partial_x) \tilde{X}_R \cr
& + \partial_x \tilde{X}_L (- \partial_t - v_L \partial_x)\tilde{X}_L + \epsilon^{\mu \nu} A_\mu \partial_\nu \Big(g_+ (X_R \cr
& + \tilde{X}_L) + g_- (X_L + \tilde{X}_R) \Big) \Big],
\end{align}
where $v_{R/L} = \sqrt{V_{11} V_{22}} \pm V_{12}$ and $g_{\pm} = \sqrt{2}({1 \over \sqrt{g}} \pm \sqrt{g})$.
Recall that the tilded and un-tilded fields are spatially separated and refer to edge modes on the bottom and top of the Hall bar, respectively.
The gauge field $A_\mu$ that multiplies these fields is understood to be evaluated at the location of the edge mode that it multiplies.

It is convenient to set all velocities $v_R = v_L = 1$.
This simplification does not affect the phase diagram of the point-contact perturbed Hall bar as the velocities of the edge modes do not affect tunneling exponents;
nor does this affect the two-terminal conductances calculated in later sections since the action in Eqn. (\ref{actionfull}) is diagonal.
(If Eqn. (\ref{actionfull}) happened not to be diagonal, a conductance would, in general, depend upon the velocities.\cite{KFPrandom})
Setting the velocities to all be equal will help simplify the formulas presented later.
We will allow arbitrary $g_{\pm}$ consistent with the assumption that all chiral edge modes move with the same velocity.

\section{Edge Equilibration}
\label{equilibration}

The Kubo formula can be used to express the two-terminal electrical Hall conductance $G$ for the edge modes on the top and bottom of a Hall bar in terms of a current-current correlation function:
\begin{align}
\label{kubo}
G = {e^2 \over \hbar \ell} \int_{0}^\ell dx' \lim_{\omega \rightarrow 0} \int_{- \infty}^\infty d \tau e^{i \omega \tau} {\langle J(x,\tau) J(x',0) \rangle \over \omega},
\end{align}
where we have analytically continued to imaginary time, $t \rightarrow i \tau$.
The conductance in Eqn. (\ref{kubo}) relates the current $J(x, \tau)$ through the point $x \in [0, \ell]$ found in linear response to a constant electric field applied to a length $\ell$ section within a wire of infinite extent.

Notice that we take the $\omega \rightarrow 0$ limit before performing the integral over $x'$.\cite{maslovstone}
In the DC $\omega \rightarrow 0$ limit, the right-hand-side of Eqn. (\ref{kubo}) does not depend upon where in $[0, \ell]$ the particular $x$ is chosen.
In Section \ref{conductance}, we will evaluate two other conductances by varying the choice of current-current correlation function considered.
Note that the conductance above is denoted in Section \ref{conductance} by $G_{{\bf \hat{x}}, {\bf \hat{x}}}$.

The total current running through the point $x$ along the top and bottom of the Hall bar,
\begin{align}
J(x, \tau) & = - {i \over 2 \pi} \partial_\tau \Big(\phi_1(x, \tau) + \phi_2(x,\tau) + \tilde{\phi}_1(x,\tau) + \tilde{\phi}_2(x,\tau)\Big) \cr
& = {g_+ \over 2 \pi} \Big(\partial_{z} \tilde{X}_L(z) - \bar{\partial}_{\bar{z}} X_R(\bar{z})\Big) \cr
& + {g_- \over 2\pi} \Big(\partial_{z} X_L(z) - \bar{\partial}_{\bar{z}} \tilde{X}_R(\bar{z})\Big),
\end{align}
where $z = x + i \tau, \bar{z} = x - i \tau, \partial_z = {1 \over 2}(\partial_x - i \partial_\tau)$, and $\bar{\partial}_{\bar{z}} = {1 \over 2}(\partial_x + i \partial_\tau)$.

We distinguish the two-terminal Hall conductance $G$ from the electrical Hall conductivity previously denoted by $\sigma_{xy}$.
$G$ generally differs from $\sigma_{xy}$ in that it depends upon properties of and conditions at the edge while $\sigma_{xy}$, as defined previously, is an (intensive) property of the bulk state.
Indeed, because the bosonic IQH state is non-chiral, the value of the conductance $G$ depends upon the $V_{IJ}$ interactions.\cite{KFPrandom, KFrandom}
Using the action, Eqn. (\ref{actionfull}), we find that Eqn. (\ref{kubo}) evaluates to $G = ({1 \over g} + g) {e^2 \over h}$.
Details of the calculation that leads to this conclusion can be found in Section \ref{conductance}.

The interaction-dependent value for the two-terminal Hall conductance is meaningful since an actual experimentalist measuring the Hall conductance runs a current along the bar, measures the voltage difference between the top and bottom of the Hall bar, and then takes the ratio of the current to voltage drop in order to abstract the Hall conductance.
This should be contrasted with the conductance of interacting 1D quantum channels for which it is necessary to attach leads to the sample which, if Fermi liquid-like, return a conductance of $e^2/h$ per channel.\cite{maslovstone, SafiSchulz} 
The use of the Kubo formula in Eqn. (\ref{kubo}) generalizes\cite{KFPrandom, KFrandom} to the interacting case the Landauer-Buttiker formalism for the study of the conductance of non-interacting electrons.\cite{Landauer, Buttiker}

Only at $g = 1$ do we obtain the equality $G = \sigma_{xy} = 2 e^2/h$.
The key physical reason for the difference between $G$ and $\sigma_{xy}$ when $g \neq 1$ is due to the lack of equilibration between modes on a given edge.
We have seen how charge conservation and translation invariance prevent mass-generating couplings between $\phi_1$ with $\phi_2$.
However, these symmetries also prevent edge equilibration.
We generally expect any physically realized edge to have some amount of disorder that breaks translation invariance.
Such disorder allows the transfer of charge between the two modes with the impurities absorbing the excess momenta.
In other words, impurities allow $\phi_1$ and $\phi_2$ to equilibrate.
  
Thus, we are lead to consider charge and momentum exchange via an impurity.
It is sufficient to consider a single edge for this analysis and so we concentrate on the top edge; an identical conclusion is drawn for the bottom edge.
The lowest dimension tunneling term between $\phi_1$ and $\phi_2$ that preserves charge conservation has the form,
\begin{align}
\label{disorder}
S_{{\rm disorder}} = \int dt dx \Big[ \xi(x) e^{i \phi_1 -i \phi_2} + {\rm h.c.} \Big],
\end{align}
where $\xi(x)$ is a complex Gaussian random variable satisfying, $\langle \langle \xi^*(x) \xi(x') \rangle \rangle = D \delta(x- x')$.
The double bracket denotes a disorder average.
The above term tunnels a $\phi_1$ mode into a $\phi_2$ mode and vice versa.
Any momentum mismatch is absorbed by the field $\xi$.

$D$ functions as a coupling constant for this interaction between $\phi_1$ and $\phi_2$.
The leading RG equation for $D$ takes the form,\cite{GiamarchiSchulz}
\begin{align}
{\partial D \over \partial \ell} = \Big(3 - 2 \Delta_R(g)\Big) D,
\end{align}
where $\Delta_R(g) = {1 \over 2} (g + {1 \over g})$ equals the scaling dimension of the operator, $\exp(i \phi_1 - i \phi_2)$.
Thus, disorder is relevant if ${1 \over 2} (3 - \sqrt{5}) < g < {1 \over 2} (3 + \sqrt{5})$.

Let us suppose that we are in a region of parameter space where disorder is relevant.
For general $g$ within this regime, perturbation theory is certainly not reliable.
To access the strong coupling fixed point, we switch to the charged $\phi_\rho$ and neutral $\phi_\sigma$ fields:
\begin{align}
\phi_\rho = & {1 \over 2 \sqrt{2}} \Big( g_+ X_R + g_- X_L \Big), \cr
\phi_\sigma = & {1 \over 2 \sqrt{2}} \Big( g_- X_R + g_+ X_L \Big).
\end{align}
The motivation for this redefinition is that it is $\phi_\rho$ that enters into the definition of the charge current in Eqn. (\ref{actionfull}) while only the neutral field $\phi_\sigma$ is involved in the disorder-mediated interaction, Eqn (\ref{disorder}).

Let us write the action in terms of these fields and show that at the strong-disorder fixed point, the charge and neutral sectors decouple.
Setting the background gauge field to zero, the action for the edge modes on the top edge in Eqn. (\ref{actionfull}) may be rewritten as,
\begin{align}
\label{chargeneutralaction}
S & = {1 \over 4 \pi} \int dt dx \Big[ \partial_x \phi_\rho(\partial_t - {v \over 2} \partial_x ) \phi_\rho \cr
& + \partial_x \phi_\sigma (- \partial_t - {v \over 2} \partial_x) \phi_\sigma - v_{\rho \sigma} \partial_x \phi_\rho \partial_x \phi_\sigma \cr
& + (\xi(x) e^{i \sqrt{2} \phi_\sigma} + {\rm h. c.}) \Big],
\end{align}
where
\begin{align}
v = & g + {1 \over g}, \cr
v_{\rho \sigma} = & g - {1 \over g}.
\end{align}
Notice the appearance of an emergent $SU(2)$ global symmetry generated by the currents $\cos(\sqrt{2} \phi_\sigma)$, $\sin(\sqrt{2} \phi_\sigma)$, and $\partial_t \phi_\sigma$ when $v_{\rho \sigma} = 0$.
This occurs when $g = 1$.
The emergent $SU(2)$ symmetry allows an exact solution for the neutral sector of the model.
(The charged sector is already described by a free field theory at this decoupled point.)
Considering small deviations away from the decoupled fixed point, the exact solution shows that the decoupled fixed point action, Eqn. (\ref{chargeneutralaction}) at $v_{\rho \sigma} = 0$, describing the charged and neutral modes is attractive.\cite{KFPrandom}

In summary, disorder, if relevant, drives the system towards the $g = 1$ point, where the two-terminal Hall conductance takes its universal value (from the bulk perspective), $G = \sigma_{xy} = 2 e^2/h$.
Interestingly, the point-contact perturbation is marginal at this fixed point as we will see shortly.

\section{Point-Contact Tunneling Fixed Points}
\label{pointcontact}

\subsection{Point-Contact Perturbations}

For well separated boundaries, terms that couple the two sets of edge modes together are absent due to locality.
The matrix element defining the possible interaction is exponentially small as long as the bulk gap is non-zero.
Indeed, a tunneling operator of lowest dimension tunnels a single boson from one edge to the other and, in general, has amplitude proportional to $\exp(- b M W)$ where $M$ is the bulk gap, $W$ is the distance between the top and bottom edges, and $b$ is a positive constant.

However, a point-contact constriction allows non-zero tunneling between the top and bottom edge modes at a single spatial location.
See Fig. 1 for the relevant geometrical pictures illustrating the possible tunneling.
About the fully connected geometry displayed in Fig. 1A, the lowest dimension tunneling operators take the form,
\begin{align}
\label{tunnelingop}
{\cal O}_{I,J}(x) & = \cos\Big(\phi_I - \tilde{\phi}_J\Big) \delta(x),
\end{align}
where we have chosen coordinates so that the single tunneling event occurs at the origin.
The operator ${\cal O}_{I,J}$ tunnels a unit charge boson from mode $\phi_I$ into mode $\tilde{\phi}_J$ at $x=0$.
From the action Eqn. (\ref{actionfull}), we determine the (boundary) scaling dimensions $\Delta_{I,J}$ of the four lowest dimension tunneling terms to be:
\begin{align}
\Delta_{I, J} = {1 \over 2} \Big(g^{2I-3} + g^{2J - 3} \Big),
\end{align}
for $I, J = 1,2$.
(Note that depending upon the values of $g$, some subset of the four operators could have higher harmonics that tunnel multiple bosons at a time which are more relevant than some of the other listed fields; when listing the four operators above, we are implicitly considering their higher harmonics as well and so we will not mention them explicitly further.)
The resulting RG equations for each boundary operator with coupling $c_{I,J}$  is then,
\begin{align}
{1 \over c_{I, J}} {\partial c_{I, J} \over \partial \ell} & = 1 - \Delta_{I, J}
\end{align}
At $g = 1$, all four operators are marginal, while for $g \neq 1$ there is a single relevant operator.
For a particular choice of $g$, the relevant operator ${\cal O}_{I,J}$ can be determined from the values of $I, J = 1,2$ satisfying $g^{3- 2I}>1$ and $g^{3-2J} >1$. 
We see that for $g \neq 1$ both ${\cal O}_{1,2}$ and ${\cal O}_{2,1}$ are irrelevant; for $g>1$, ${\cal O}_{1,1}$ is relevant, while for $g<1$, ${\cal O}_{2,2}$ is relevant.

In the previous section, we observed that edge equilibration is relevant when ${1 \over 2} (3 - \sqrt{5}) < g < {1 \over 2} (3 + \sqrt{5})$, and the system is driven towards a strong-disorder fixed point where $g = 1$ and all four tunneling terms are marginal.
The point-contact analysis in the remainder of this section assumes either $g$ is outside the attractive regime of this strong-disorder fixed point, or that the edge is sufficiently clean and studied at high enough temperatures so that disorder has not had ``time" to renormalize the system to the $g = 1$ point.

We emphasize that for $g \neq 1$, either ${\cal O}_{1,1}$
or ${\cal O}_{2,2}$ is relevant at weak coupling, but not both. 
It is this fact that results in an easily tractable IR attractive fixed point describing a partially-split Hall droplet. 
The single relevant operator drives a (boundary) RG trajectory where the Hall bar partially
pinches off; however, it does not split into two disconnected liquids. 
We schematically illustrate the partially-split geometry corresponding to this fixed point in Fig. 1B

In the next two sections, we describe the boundary fixed point that results from perturbation by ${\cal O}_{1,1}$ (${\cal O}_{2,2}$) for $g>1$ ($g<1$) and demonstrate its stability.
(In fact, these two fixed points can be mapped into one another by a simple change of variables.)
To this end, consider the fully disconnected geometry drawn in Fig. 1C where there are two disconnected Hall bars coupled together at a single point contact.
As we will describe, from precisely the same analysis as above, there exists a single relevant (boundary) tunneling term that drives an RG trajectory towards a partially-split geometry.
We will provide evidence that this IR fixed point is the same as the one obtained by starting from the fully connected Hall droplet of Fig. 1A, thus implying a single, isolated IR stable fixed point separating the fixed points describing the fully connected and fully disconnected geometries.

We caution that our analysis is valid at weak initial tunneling coupling.
Our results assume that the single relevant tunneling term dominates the IR physics.
This assumption is certainly reasonable given our setup and we confirm its consistency in our analysis below.
However, we point out that when $V_{IJ} \neq \tilde{V}_{IJ}$, there exist parameter regimes where two relevant tunneling terms compete (there is always at least one relevant operator).
We leave the interesting question of the precise characterization of the resulting IR fixed point(s) to future work.

\subsection{Repulsive $g>1$ Interactions}

\subsubsection{The Approach from the Fully Connected Fixed Point}

When $g>1$, ${\cal O}_{1,1} = \cos(\phi_1 - \tilde{\phi}_1) \delta(x)$ is the most relevant tunneling operator.
It has (boundary) scaling dimension $\Delta_{1,1} = {1 \over g}$.
Tunneling of the $\phi_2$ boson into the $\tilde{\phi}_2$ boson by the operator ${\cal O}_{2,2}$ is strictly irrelevant.

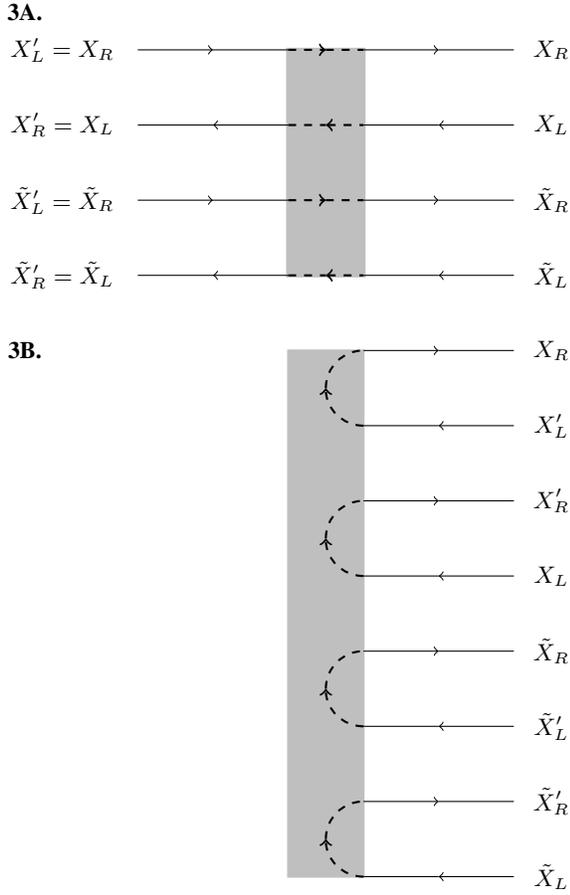
\begin{figure}[t]
\begin{tikzpicture}
\draw [fill=red, ultra thick,lightgray] (2,0) rectangle (3,-3);
\node at (-1.5,.5) {{\bf 3A.}};
\node at (-1,0) {$X'_L = X_R$};
\draw [->] (0,0) -- (1,0);
\draw (1,0) -- (2,0);
\draw [black] [thick, dashed] [->] (2,0) -- (2.5,0);
\draw [black] [thick, dashed] (2.5,0) -- (3,0);
\draw [->] (3,0) -- (4,0);
\draw (4,0) --(5,0);
\node at (5.5,0) {$X_R$};
\node at (-1,-1) {$X'_R = X_L$};
\draw (0,-1) -- (1,-1);
\draw [<-] (1,-1) -- (2,-1);
\draw [black] [thick, dashed] (2, -1) -- (2.5, -1);
\draw [<-] [black] [thick, dashed] (2.5, -1) -- (3,-1);
\draw (3,-1) -- (4,-1);
\draw [<-] (4,-1) --(5,-1);
\node at (5.5,-1) {$X_L$};
\node at (-1,-2) {$\tilde{X}'_L = \tilde{X}_R$};
\draw [->] (0,-2) -- (1,-2);
\draw (1,-2) -- (2,-2);
\draw [black] [thick, dashed] [->] (2,-2) -- (2.5,-2);
\draw [black] [thick, dashed] (2.5,-2) -- (3,-2);
\draw [->] (3,-2) -- (4,-2);
\draw (4,-2) --(5,-2);
\node at (5.5,-2) {$\tilde{X}_R$};
\node at (-1,-3) {$\tilde{X}'_R = \tilde{X}_L$};
\draw (0,-3) -- (1,-3);
\draw [<-] (1,-3) -- (2,-3);
\draw [black] [thick, dashed] (2, -3) -- (2.5, -3);
\draw [<-] [black] [thick, dashed] (2.5, -3) -- (3,-3);
\draw (3,-3) -- (4,-3);
\draw [<-] (4,-3) --(5,-3);
\node at (5.5,-3) {$\tilde{X}_L$};
\node at (-1.5,-4) {{\bf 3B.}};
\draw [fill=red, thick, lightgray] (2,-4) rectangle (3,-11);
\draw [->] [black] [thick, dashed] (3,-5) arc [radius=.5, start angle=270, end angle= 180];
\draw [black] [thick, dashed] (2.5,-4.5) arc [radius=.5, start angle=180, end angle= 90];
\draw [->] (3, -4) -- (4, -4);
\draw (4, -4) -- (5, -4);
\node at (5.5, -4) {$X_R$};
\draw (3, -5) -- (4, -5);
\draw [<-] (4, -5) -- (5, -5);
\node at (5.5, -5) {$X'_L$};
\draw [->] [black] [thick, dashed] (3,-7) arc [radius=.5, start angle=270, end angle= 180];
\draw [black] [thick, dashed] (2.5,-6.5) arc [radius=.5, start angle=180, end angle= 90];
\draw [->] (3, -6) -- (4, -6);
\draw (4, -6) -- (5, -6);
\node at (5.5, -6) {$X'_R$};
\draw (3, -7) -- (4, -7);
\draw [<-] (4, -7) -- (5, -7);
\node at (5.5, -7) {$X_L$};
\draw [->] [black] [thick, dashed] (3,-9) arc [radius=.5, start angle=270, end angle= 180];
\draw [black] [thick, dashed] (2.5,-8.5) arc [radius=.5, start angle=180, end angle= 90];
\draw [->] (3, -8) -- (4, -8);
\draw (4, -8) -- (5, -8);
\node at (5.5, -8) {$\tilde{X}_R$};
\draw (3, -9) -- (4, -9);
\draw [<-] (4, -9) -- (5, -9);
\node at (5.5, -9) {$\tilde{X}'_L$};
\draw [->] [black] [thick, dashed] (3,-11) arc [radius=.5, start angle=270, end angle= 180];
\draw [black] [thick, dashed] (2.5,-10.5) arc [radius=.5, start angle=180, end angle= 90];
\draw [->] (3, -10) -- (4, -10);
\draw (4, -10) -- (5, -10);
\node at (5.5, -10) {$\tilde{X}'_R$};
\draw (3, -11) -- (4, -11);
\draw [<-](4, -11) -- (5, -11);
\node at (5.5, -11) {$\tilde{X}_L$};
\end{tikzpicture}
\caption{The Folding Procedure.
The point contact is represented by the shaded rectangle and the boundary conditions at the fully connected fixed point are shown using dotted arrows.
In Fig. 3A, we draw the un-folded geometry; the fields on the left of the point contact are renamed as indicated.
In Fig. 3B, we draw the folded geometry.}
\label{figthree}
\end{figure}

To understand the resulting boundary fixed point obtained after perturbation by ${\cal O}_{1,1}$ in Regions I-III, we ``fold" the geometry about the location of the point contact.\cite{wongafflecktunnel}
Our folding conventions are shown in Fig. 3.
The folding procedure is merely a convenient method for analyzing the problem that makes clear the structure of the resulting boundary fixed point.

To this end, the Luttinger liquid extends along the x-axis from $-L$ to $+L$ with the understanding that $L \rightarrow \infty$.
The point contact will be placed at $x = 0$.
We fold by defining the fields,
\begin{align}
\label{folding}
X_R(x) & = X_R(x),\quad  X'_L(x)  =  X_R(- x), \quad x \in [0, L], \cr
X_L(x) & = X_L(x),\quad  X'_R(x)  = X_L(- x), \quad x \in [0, L], \cr
\tilde{X}_R(x) & = \tilde{X}_R(x), \quad \tilde{X}'_L(x)  =  \tilde{X}_R(- x), \quad x \in [0, L], \cr
\tilde{X}_L(x) & = \tilde{X}_L(x), \quad \tilde{X}'_R(x)  = \tilde{X}_L(- x), \quad x \in [0, L].
\end{align}
The folding operation has introduced the primed fields which are simply the continuation to $x \in [-L, 0]$ of the un-primed fields.
Note that a right/left moving field on the negative x-axis becomes a left/right moving field when redefined to live on the positive x-axis.

When there is no point contact present, the field redefinition in Eqn. (\ref{folding}) is unnecessary.
However, it will prove very useful in understanding the nature of the partially-split fixed point and for computing current-current correlation functions in the next section.

It is important to compare the boundary conditions obeyed by the fields in the fully connected geometry to the boundary conditions obtained at the putative IR fixed point induced by ${\cal O}_{1,1}$.
When there is no point contact, the boundary conditions on the fields at $x=0$ are simply:
\begin{align}
\label{fulltrans}
\phi_1(0) & = \phi'_1(0),\quad \phi_2(0) = \phi'_2(0), \cr
\tilde{\phi}_1(0) & = \tilde{\phi}'_1(0), \quad \tilde{\phi}_2(0) = \tilde{\phi}'_2(0),
\end{align}
where $\phi'_1 = {1 \over \sqrt{2 g}}(X'_L + X'_R), \phi'_2 = {\sqrt{g} \over \sqrt{2}}(X'_L - X'_R), \tilde{\phi}'_1 = {1 \over \sqrt{2 g}}(\tilde{X}'_R + \tilde{X}'_L)$, and $\tilde{\phi}'_2 = {\sqrt{g} \over \sqrt{2}}(- \tilde{X}'_L + \tilde{X}'_R)$.
We will refer to the boundary conditions when there is no point contact as the fully connected boundary conditions with the corresponding geometry shown in Fig. 1A.
In the folded setup, these boundary conditions are displayed in Fig. 3B.

In order to study the non-trivial boundary fixed point induced by the perturbation ${\cal O}_{1,1}$, it is convenient to first make the following field redefinitions by introducing the right and left moving fields $R_i$ and $L_j$ for $i,j = 1, ..., 4$:
\begin{widetext}
\begin{alignat}{2}
\label{leftrightdef}
X_R & ={1 \over 2} (R_1 + R_2 + R_3 + R_4), & \quad X'_R & = {1 \over 2} (- R_1 - R_2 + R_3 + R_4), \cr
\tilde{X}_R & = {1 \over 2} (R_1 - R_2 + R_3 - R_4), & \quad \tilde{X}'_R & = {1 \over 2} (- R_1 + R_2 + R_3 - R_4), \cr
X_L & = {1 \over 2} (- L_1 -L_2 + L_3 + L_4), & \quad X'_L & = {1 \over 2} (L_1 + L_2 + L_3 + L_4), \cr
\tilde{X}_L & = {1 \over 2} (- L_1 + L_2 + L_3 - L_4), & \quad \tilde{X}'_L & = {1 \over 2} (L_1 - L_2 + L_3 - L_4).
\end{alignat}
\end{widetext}
All fields are understood to live on the half-line, $x \in [0, L]$.
For convenience, we provide the inverse of the transformation in Eqn. (\ref{leftrightdef}):
\begin{widetext}
\begin{alignat}{2}
\label{inversetrans}
R_1 & = {1 \over 2} (X_R - X'_R + \tilde{X}_R - \tilde{X}'_R), & \quad R_2 & = {1 \over 2} (X_R - X'_R - \tilde{X}_R + \tilde{X}'_R), \cr
R_3 & = {1 \over 2} (X_R + X'_R + \tilde{X}_R + \tilde{X}'_R), & \quad R_4 & = {1 \over 2} (X_R + X'_R - \tilde{X}_R - \tilde{X}'_R), \cr
L_1 & = {1 \over 2} (- X_L + X'_L - \tilde{X}_L + \tilde{X}'_L), & \quad L_2 & = {1 \over 2} (- X_L + X'_L + \tilde{X}_L - \tilde{X}'_L), \cr
L_3 & = {1 \over 2} (X_L + X'_L + \tilde{X}_L + \tilde{X}'_L), & \quad L_4 & = {1 \over 2} (X_L + X'_L - \tilde{X}_L - \tilde{X}'_L).
\end{alignat}
\end{widetext}
The action remains diagonal after the field redefinitions in Eqn. (\ref{leftrightdef}), however, the coupling to the gauge field is changed.
In terms of these fields, the fully connected boundary conditions in Eqn. (\ref{fulltrans}) become:
\begin{align}
\label{fulltranstwo}
R_i(0) = L_i(0),
\end{align}
for all $i$.

We are now ready to study the effects of the point contact.
An important point is that because we have (redundantly) doubled the number of fields in folding the geometry, a point contact now corresponds to two boundary perturbations.
In the folded geometry, the operator ${\cal O}_{1,1}$ becomes:
\begin{align}
{\cal O}_{1,1} \rightarrow \cos(\phi_1 - \tilde{\phi}_1) \delta(x) + \cos(\phi'_1 - \tilde{\phi}'_1) \delta(x).
\end{align}
Hopefully without leading to any confusion, we will continue to refer to this operator in the folded geometry as ${\cal O}_{1,1}$.

The benefit of the field redefinitions in Eqn. (\ref{leftrightdef}) is that the boundary conditions induced by the operator ${\cal O}_{1,1}$ are easy to analyze.
The effect of the point-contact perturbation, ${\cal O}_{1,1}$, is to freeze:
\begin{align}
\label{pointconstraint}
\phi_1(0) = & \tilde{\phi}_1(0), \cr
\phi'_1(0) = & \tilde{\phi}'_1(0).
\end{align}
These two conditions take a simple form when written in terms of the $R_i, L_j$ fields: $R_2(0) = L_2(0)$ and $R_4(0) = - L_4(0)$.
Thus, the point contact only changes the boundary condition relating $R_4$ to $L_4$.
The boundary conditions on the other pair of fields, $R_i, L_j$ for $i,j=1,3$, are unaffected by the point contact and, therefore, remain the same as in the fully connected case, Eqn. (\ref{fulltranstwo}).

Summarizing, we find that the point-contact perturbation drives a boundary RG flow to the fixed point characterized by the boundary conditions,
\begin{align}
\label{partialtransone}
R_i(0) & = L_i(0),\ i = 1,2,3, \cr
R_4(0) & = - L_4(0).
\end{align}
If we form the non-chiral bosons $\chi_i = R_i + L_i$, then the effect of the point contact is to drive the (integrable) boundary flow from the Neumann to the Dirichlet boundary condition for $\chi_4$ with the other three fields maintaining their initial Neumann conditions at the point contact.
The boundary conditions in Eqn. (\ref{partialtransone}) define the partially-split fixed point and, for brevity, we shall refer to these boundary conditions as the partially-split boundary conditions.

\subsubsection{The Approach from the Fully Disconnected Fixed Point}

\begin{figure}[t]
\begin{tikzpicture}
\draw [fill=red, ultra thick,lightgray] (-.2,0) rectangle (3,-3);
\node at (-2.7,.5) {{\bf 4A.}};
\node at (-2.7,0) {$X'_L$};
\draw [->] (-2.2,0) -- (-1.2,0);
\draw (-1.2,0) -- (-.2,0);
\draw [->][black] [thick, dashed] (3,-3) arc [radius=1.5, start angle=270, end angle= 180];
\draw [black] [thick, dashed] (1.5,-1.5) arc [radius=1.5, start angle=180, end angle= 90];
\draw [<-] [black] [thick, dashed] (1.3,-1.5) arc [radius=1.5, start angle=0, end angle= 90];
\draw [black] [thick, dashed] (-.2,-3) arc [radius=1.5, start angle=-90, end angle= 0];
\draw [->] (3,0) -- (4,0);
\draw (4,0) --(5,0);
\node at (5.5,0) {$X_R$};
\node at (-2.7,-1) {$X'_R$};
\draw (-2.2,-1) -- (-1.2,-1);
\draw [<-] (-1.2,-1) -- (-.2,-1);
\draw [black] [thick, dashed] (3,-2) arc [radius=.5, start angle=270, end angle= 180];
\draw [<-] [black] [thick, dashed] (2.5,-1.5) arc [radius=.5, start angle=180, end angle= 90];
\draw [black] [thick, dashed] (.3,-1.5) arc [radius=.5, start angle=0, end angle= 90];
\draw [->] [black] [thick, dashed] (-.2,-2) arc [radius=.5, start angle=-90, end angle= 0];
\draw (3,-1) -- (4,-1);
\draw [<-] (4,-1) --(5,-1);
\node at (5.5,-1) {$X_L$};
\node at (-2.7,-2) {$\tilde{X}'_L$};
\draw [->] (-2.2,-2) -- (-1.2,-2);
\draw (-1.2,-2) -- (-.2,-2);
\draw [->] (3,-2) -- (4,-2);
\draw (4,-2) --(5,-2);
\node at (5.5,-2) {$\tilde{X}_R$};
\node at (-2.7,-3) {$\tilde{X}'_R$};
\draw (-2.2,-3) -- (-1.2,-3);
\draw [<-] (-1.2,-3) -- (-.2,-3);
\draw (3,-3) -- (4,-3);
\draw [<-] (4,-3) --(5,-3);
\node at (5.5,-3) {$\tilde{X}_L$};
\node at (-2.7,-3.5) {{\bf 4B.}};
\draw [fill=red, thick, lightgray] (-1,-4) rectangle (3,-11);
\draw [black] [thick, dashed] (3,-8) arc [radius=.5, start angle=270, end angle= 180];
\draw [<-] [black] [thick, dashed] (2.5,-7.5) arc [radius=.5, start angle=180, end angle= 90];
\draw [->] (3, -4) -- (4, -4);
\draw (4, -4) -- (5, -4);
\node at (5.5, -4) {$X_R$};
\draw (3, -5) -- (4, -5);
\draw [<-] (4, -5) -- (5, -5);
\node at (5.5, -5) {$X'_L$};
\draw [->] [black] [thick, dashed] (3,-9) arc [radius=1.5, start angle=270, end angle= 180];
\draw [black] [thick, dashed] (1.5,-7.5) arc [radius=1.5, start angle=180, end angle= 90];
\draw [->] (3, -6) -- (4, -6);
\draw (4, -6) -- (5, -6);
\node at (5.5, -6) {$X'_R$};
\draw (3, -7) -- (4, -7);
\draw [<-](4, -7) -- (5, -7);
\node at (5.5, -7) {$X_L$};
\draw [black] [thick, dashed] (3,-10) arc [radius=2.5, start angle=270, end angle= 180];
\draw [<-] [black] [thick, dashed] (.5,-7.5) arc [radius=2.5, start angle=180, end angle= 90];
\draw [->] (3, -8) -- (4, -8);
\draw (4, -8) -- (5, -8);
\node at (5.5, -8) {$\tilde{X}_R$};
\draw (3, -9) -- (4, -9);
\draw [<-] (4, -9) -- (5, -9);
\node at (5.5, -9) {$\tilde{X}'_L$};
\draw [->] [black] [thick, dashed] (3,-11) arc [radius=3.5, start angle=270, end angle= 180];
\draw [black] [thick, dashed] (-.5,-7.5) arc [radius=3.5, start angle=180, end angle= 90];
\draw [->] (3, -10) -- (4, -10);
\draw (4, -10) -- (5, -10);
\node at (5.5, -10) {$\tilde{X}'_R$};
\draw (3, -11) -- (4, -11);
\draw [<-](4, -11) -- (5, -11);
\node at (5.5, -11) {$\tilde{X}_L$};
\end{tikzpicture}
\caption{The boundary conditions at the fully disconnected fixed point.
The point contact is represented by the shaded rectangle and the boundary conditions at the fully disconnected fixed point are shown using dotted arrows.
Figs. 4A and 4B show the respective un-folded and folded geometries.
(Compared to Fig. 3, we have smeared out the point contact even further in order to avoid line crossings in the figure.)}
\label{figfour}
\end{figure}
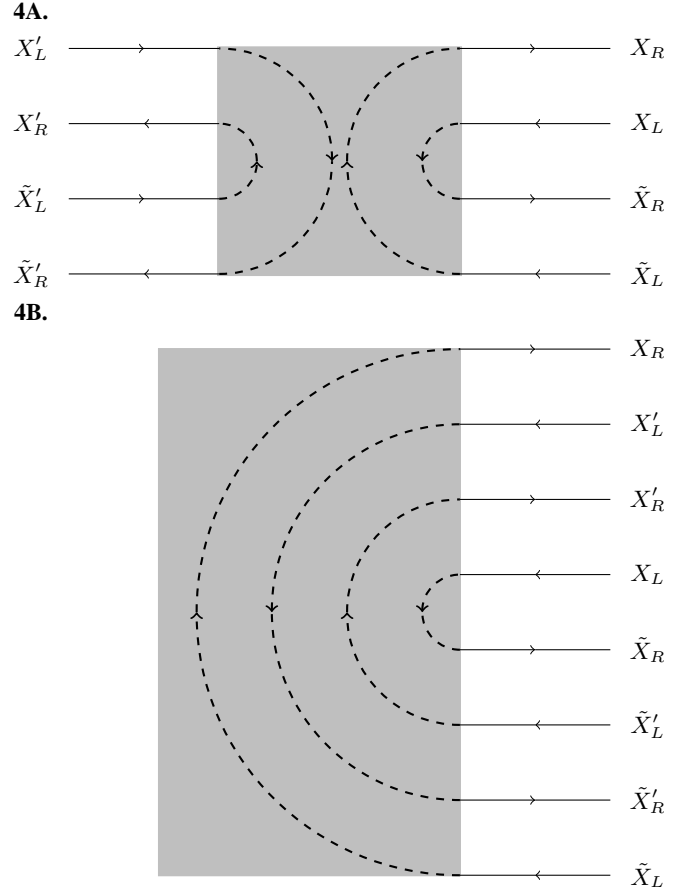

We now wish to give evidence for the stability or attractiveness of the partially-split fixed point.
As we have seen, starting from the fully connected fixed point, perturbation by the relevant operator ${\cal O}_{1,1}$ drives the system towards the partially-split fixed point.

Consider instead the approach to the partially-split fixed point from the fully disconnected geometry where there are two bosonic IQH droplets interacting via a single point contact drawn schematically in Figure 1C.
In terms of the folded fields of Fig. 3A, the fully disconnected geometry is defined by the following boundary conditions:
\begin{align}
\label{disconnected}
\phi_1(0) & = \tilde{\phi}_1(0), \quad \phi_2(0) = \tilde{\phi}_2(0), \cr
\phi'_1(0) & = \tilde{\phi}'_1(0), \quad \phi'_2(0) = \tilde{\phi}'_2(0). 
\end{align}
To facilitate comparison with the fully connected boundary conditions in Fig. 3, the fully disconnected boundary conditions are drawn in Fig. 4.
Rewritten in terms of the $R_i, L_j$ fields, these conditions become:
\begin{alignat}{2}
\label{disconnectedother}
R_1(0) & = - L_1(0), & \quad R_2(0) & = L_2(0),\cr
R_3(0) & = L_3(0), & \quad R_4(0) & = - L_4(0).
\end{alignat}
Thus, if the partially-split fixed point is an attractor, the effect of the point contact at the fully disconnected fixed point must be to change the boundary condition relating $R_1$ and $L_1$ in order to match Eqn. (\ref{partialtransone}).

To verify that this does indeed occur, we shall un-fold the geometry beginning at the disconnected fixed point boundary conditions, Eqn. (\ref{disconnected}), and show that the leading point-contact perturbation drives the theory to the partially-split fixed point.
(The following un-folding and re-folding is not necessary for this analysis. 
Instead, we may arrive at the same conclusion by analyzing the effects of the leading point-contact perturbation at the fixed point defined in Eqn. (\ref{disconnected}).)
Using Eqn. (\ref{disconnected}) and Fig. \ref{figfour}, we define the fields
\begin{align}
\label{leftrightchiralsdisconnect}
\varphi_1 & = {1 \over \sqrt{2 g}}(X_R + X_L), \quad \varphi_2 = {\sqrt{g} \over \sqrt{2}} (X_R - X_L),\cr
\tilde{\varphi}_1 & = {1 \over \sqrt{2 g}}(X'_R + X'_L), \quad \tilde{\varphi}_2 = {\sqrt{g} \over \sqrt{2}}(- X'_R + X'_L).
\end{align}
In making these definitions, we have identified $\tilde{X}_L(x) = X_R(-x)$ and similarly for the other tilded fields shown in Fig. \ref{figfour}.
The action for the fully disconnected geometry takes exactly the same form, Eqn. (\ref{startaction}), as the action for the fully connected geometry with the replacements $\phi_I \leftrightarrow \varphi_I$, $\tilde{\phi}_I \leftrightarrow \tilde{\varphi}_I$ (along with the understanding that we are now describing edge modes belonging to disconnected Hall samples).

The leading point-contact perturbation in the un-folded variables takes the form: ${\cal O}'_{1,1} = \cos(\varphi_1 - \tilde{\varphi}_1) \delta(x)$.
To analyze the effects of this operator, we fold the geometry as shown in Fig. \ref{figfour}.
(In contrast to how we folded when starting at the fully connected fixed point, we now introduce tilded left and right moving fields instead of primed left and right moving fields.)
As we previously noted, folding splits the point-contact perturbation into two boundary operators and imposes the constraints:
\begin{align}
\label{disconnectpcconstraint}
X_R + X_L & = X'_R + X'_L, \cr
\tilde{X}_R + \tilde{X}_L & = \tilde{X}'_R + \tilde{X}'_L.
\end{align}
By taking linear combinations of these two equations and using Eqns. (\ref{inversetrans}), we see that in terms of the $R_i$ and $L_j$ variables, these boundary conditions become: $R_1 = L_1$ and $R_2 = L_2$.
${\cal O}'_{1,1}$ does not affect the boundary conditions of $R_{3,4}$ and $L_{3,4}$ given in Eqn. (\ref{disconnectedother}).
We recognize the resulting boundary conditions as describing the partially-split fixed point.
Thus, the leading point-contact perturbation drives an RG trajectory from either the fully connected or fully disconnected fixed points to the partially-split fixed point in the IR, thereby implying the existence of a single, isolated fixed point in between these two limits.

\subsubsection{Stability of the Partially-Split Fixed Point}

Having demonstrated the symmetry between the two approaches to the partially-split fixed point, we now ask whether there exist additional instabilities at this IR fixed point.
In other words, are there potential runaway ``directions" in coupling constant space?

To determine the stability of the partially-split fixed point, we need only enumerate all boundary perturbations at the partially-split fixed point and compute their scaling dimensions.
In the unfolded language, an arbitrary tunneling perturbation takes the form:
\begin{align}
\label{ops}
{\cal O}_{a_i, b_j} = \cos\Big(a_1 \phi_1 + a_2 \phi_2 + b_1 \tilde{\phi}_1 + b_2 \tilde{\phi}_2\Big) \delta(x),
\end{align}
for $a_i, b_j \in \mathbb{Z}$.
In this notation, the four operators considered previously are written as ${\cal O}_{1,1} = {\cal O}_{(1,0), (1,0)}, {\cal O}_{1,2} = {\cal O}_{(1,0),(0,1)}, {\cal O}_{2,1} = {\cal O}_{(0,1), (1,0)}$, and ${\cal O}_{2,2} = {\cal O}_{(0,1),(0,1)}$.
We must compute the dimensions of these operators at the partially-split fixed point.
The IR scaling dimensions will generally differ from the values taken at the UV fixed point.

To begin, we need to constrain the form of the operator so that it conserves $U(1)$ charge.
Neutrality requires: $a_1 + a_2 + b_1 + b_2 = 0$.
To compute the scaling dimension, we first express ${\cal O}_{a_i, b_j}$ in terms of the $R_i, L_j$ fields.
We then impose the partially-split boundary conditions in Eqn. (\ref{partialtransone}) to write the $L_j$ fields in terms of the $R_i$.
The scaling dimensions are then read off from the temporal decay at $x=0$ of the two-point correlation function of ${\cal O}_{a_i, b_j}(t)$ finding,
\begin{align}
\label{scalingatfp}
\Delta_{a_i, b_j}(g) = {1 \over 4} \Big({(a_1 + b_1)^2 \over g} + g(3 a_2^2 + 3 b_2^2 + 2 a_2 b_2)\Big),
\end{align}
subject to the neutrality constraint. 

No operator ${\cal O}_{a_i, b_j}$ is relevant at the partially-split fixed point.
Consider the following low-dimensional examples.
${\cal O}_{1,2} = \cos(\phi_1 - \tilde{\phi}_2) \delta(x)$ and ${\cal O}_{2,1} = \cos(\tilde{\phi}_1 - \phi_2) \delta(x)$ are both of scaling dimension, $\Delta_{1,2} = \Delta_{2,1} = {1 \over 4}(3g + 1/g)$.
${\cal O}_{2,2} = \cos(\phi_2 - \tilde{\phi}_2) \delta(x)$ has scaling dimension, $\Delta_{2,2} = g$.
Both sets of operators, along with all other operators given in Eqn. (\ref{ops}), are irrelevant when $g>1$.
(The operator ${\cal O}_{1,1}$ becomes the identity operator at the partially-split fixed point as its constraint is satisfied on all states.)

\subsection{Attractive $g<1$ Interactions}

It is straightforward to apply the method outlined in the previous section to the case when $g<1$.
Therefore, we can more or less state the following results instead of providing a full derivation as in the previous subsection.

When $g<1$, the most relevant operators at the fully connected and fully disconnected fixed point are ${\cal O}_{2,2}$ and ${\cal O}'_{2,2}$, respectively (using notation introduced previously).
Just as the boundary conditions induced by ${\cal O}_{1,1}$ only involved $R_{2,4}$ and $L_{2,4}$, the conditions imposed by ${\cal O}_{2,2}$ only involve $R_{1,3}$ and $L_{1,3}$.
Around the fully connected fixed point, the conditions, $\phi^2 = \tilde{\phi}^2$ and $\phi'^2 = \tilde{\phi}'^2$ imposed by the leading point-contact perturbation when $g<1$ results in a fixed point characterized by the boundary conditions:
\begin{alignat}{2}
\label{gsmallfp}
R_1(0) & = - L_1(0), & \quad R_2(0) & = L_2(0), \cr
R_3(0) & = L_3(0), & \quad R_4(0) & = L_4(0).
\end{alignat}
(About the fully disconnected fixed point, ${\cal O}'_{2,2}$ drives the system to the same IR stable partially-split fixed point, but instead involves the fields $R_{3,4}$ and $L_{3,4}$.)
The RG diagram for this flow takes the form shown in Fig. 1 (iii).
The stability of the fixed point, Eqn. (\ref{gsmallfp}), is also immediate since the lowest dimension tunneling operator is ${\cal O}_{1,1}$ which has boundary scaling dimension equal to $1/g > 1$ for $g<1$.

Thus, the transformation $g \leftrightarrow 1/g$ and $R_1, L_1 \leftrightarrow R_4, L_4$ takes us between the fixed point induced by ${\cal O}_{1,1}$ (${\cal O}'_{1,1}$) and ${\cal O}_{2,2}$ (${\cal O}'_{2,2}$).

\section{Two-Terminal Conductances}
\label{conductance}

We now describe how the boundary conditions characterizing the various fixed points studied in the previous section are reflected in the current-current correlation functions that determine two-terminal electrical conductances at zero temperature.
The leading finite temperature corrections to the zero temperature conductances calculated here scale as $- a T^{2 \Delta - 2}$, where $\Delta$ is the smallest scaling dimension of a tunneling operator at a given fixed point and $a$ is some non-negative constant.

\begin{figure}[t]
\begin{tikzpicture}
\node at (-9, 0) {{\bf 5A.}};
\node at (-7.9, -.25) {{\tiny $X'_L$}};
\node at (-8.45, -.25) {{\tiny $X'_R$}};
\node at (-6.1, -.25) {{\tiny $X_R$}};
\node at (-5.6, -.25) {{\tiny $X_L$}};
\node at (-7.9, -2.85) {{\tiny $\tilde{X}'_R$}};
\node at (-8.45, -2.85) {{\tiny $\tilde{X}'_L$}};
\node at (-6.1, -2.85) {{\tiny $\tilde{X}_L$}};
\node at (-5.6, -2.85) {{\tiny $\tilde{X}_R$}};
\draw [->] (-8,-.5) -- (-7.5,-1);
\draw (-7.5,-1) -- (-7,-1.5);
\draw [->] (-7,-1.5) -- (-6.5, -1);
\draw (-6.5, -1) -- (-6,-.5);
\draw (-8.25, -.5) -- (-7.5, -1.25);
\draw [<-] (-7.5, -1.25) -- (-7, -1.75);
\draw (-7,-1.75) -- (-6.5, -1.25);
\draw [<-] (-6.5, -1.25) -- (-5.75, -.5);
\draw [->] (-8.25, -2.6) -- (-7.5, -1.85);
\draw (-7.5, -1.85) -- (-7, -1.35);
\draw [->] (-7,-1.35) -- (-6.5, -1.85);
\draw (-6.5, -1.85) -- (-5.85, -2.5);
\draw (-8, -2.6) -- (-7.5, -2.1);
\draw [<-] (-7.5, -2.1) -- (-7,-1.6);
\draw  (-7,-1.6) -- (-6.5, -2.1);
\draw [<-] (-6.6, -2) -- (-6,-2.6);
\draw [thick, dashed] (-6.25, -.5) -- (-6.25, -2.6);
\draw [fill=red, ultra thick,green] (-7.30,-1.25) rectangle (-6.75,-1.80);
\node at (-4, 0) {{\bf 5B.}};
\node at (-2.9, -.25) {{\tiny $X'_L$}};
\node at (-3.45, -.25) {{\tiny $X'_R$}};
\node at (-1.1, -.25) {{\tiny $X_R$}};
\node at (-0.6, -.25) {{\tiny $X_L$}};
\node at (-2.9, -2.85) {{\tiny $\tilde{X}'_R$}};
\node at (-3.45, -2.85) {{\tiny $\tilde{X}'_L$}};
\node at (-1.1, -2.85) {{\tiny $\tilde{X}_L$}};
\node at (-0.6, -2.85) {{\tiny $\tilde{X}_R$}};
\draw [->] (-3,-.5) -- (-2.5,-1);
\draw (-2.5,-1) -- (-2,-1.5);
\draw [->] (-2,-1.5) -- (-1.5, -1);
\draw (-1.5, -1) -- (-1,-.5);
\draw (-3.25, -.5) -- (-2.5, -1.25);
\draw [<-] (-2.5, -1.25) -- (-2, -1.75);
\draw (-2,-1.75) -- (-1.5, -1.25);
\draw [<-] (-1.5, -1.25) -- (-0.75, -.5);
\draw [->] (-3.25, -2.6) -- (-2.5, -1.85);
\draw (-2.5, -1.85) -- (-2, -1.35);
\draw [->] (-2,-1.35) -- (-1.5, -1.85);
\draw (-1.5, -1.85) -- (-0.85, -2.5);
\draw (-3, -2.6) -- (-2.5, -2.1);
\draw [<-] (-2.5, -2.1) -- (-2,-1.6);
\draw  (-2,-1.6) -- (-1.5, -2.1);
\draw [<-] (-1.6, -2) -- (-1,-2.6);
\draw [thick, dashed] (-3.25,-.8) -- (-.75,-.8);
\draw [fill=red, ultra thick,green] (-2.30,-1.25) rectangle (-1.75,-1.80);
\node at (-9, -4) {{\bf 5C.}};
\node at (-7.9, -4.25) {{\tiny $X'_L$}};
\node at (-8.45, -4.25) {{\tiny $X'_R$}};
\node at (-6.1, -4.25) {{\tiny $X_R$}};
\node at (-5.6, -4.25) {{\tiny $X_L$}};
\node at (-7.9, -6.85) {{\tiny $\tilde{X}'_R$}};
\node at (-8.45, -6.85) {{\tiny $\tilde{X}'_L$}};
\node at (-6.1, -6.85) {{\tiny $\tilde{X}_L$}};
\node at (-5.6, -6.85) {{\tiny $\tilde{X}_R$}};
\draw [->] (-8,-4.5) -- (-7.5,-5);
\draw (-7.5,-5) -- (-7,-5.5);
\draw [->] (-7,-5.5) -- (-6.5, -5);
\draw (-6.5, -5) -- (-6,-4.5);
\draw (-8.25, -4.5) -- (-7.5, -5.25);
\draw [<-] (-7.5, -5.25) -- (-7, -5.75);
\draw (-7,-5.75) -- (-6.5, -5.25);
\draw [<-] (-6.5, -5.25) -- (-5.75, -4.5);
\draw [->] (-8.25, -6.6) -- (-7.5, -5.85);
\draw (-7.5, -5.85) -- (-7, -5.35);
\draw [->] (-7,-5.35) -- (-6.5, -5.85);
\draw (-6.5, -5.85) -- (-5.85, -6.5);
\draw (-8, -6.6) -- (-7.5, -6.1);
\draw [<-] (-7.5, -6.1) -- (-7,-5.6);
\draw  (-7,-5.6) -- (-6.5, -6.1);
\draw [<-] (-6.6, -6) -- (-6,-6.6);
\draw [thick, dashed] (-6.25, -5.75) -- (-6.25, -6.6);
\draw [thick, dashed] (-8.25, -4.8) -- (-7.45, -4.8);
\draw [fill=red, ultra thick,green] (-7.30,-5.25) rectangle (-6.75,-5.80);
\end{tikzpicture}
\caption{The dashed lines in the above three figures denote the line through which the three currents in Eqn. (\ref{twotermcurrents}) flow; the conductances in Eqn. (\ref{conductances}) measure the flow of charge through this line.
As a point of reference, Fig. 5A measures the conductance along the Hall bar for the current $J_{{\bf \hat{x}}}$, Fig. 5B measures the conductance across the Hall bar for the current $J_{{\bf \hat{y}}}$, and Fig. 5C measures a skew conductance with respect to the fully connected fixed point of Fig. 1A for the current $J_{{\rm s}}$.
The green square represents a possible (boundary) interaction induced by a point contact and the modes are labelled according to the conventions of Fig. \ref{figthree}.}
\label{figfive}
\end{figure}
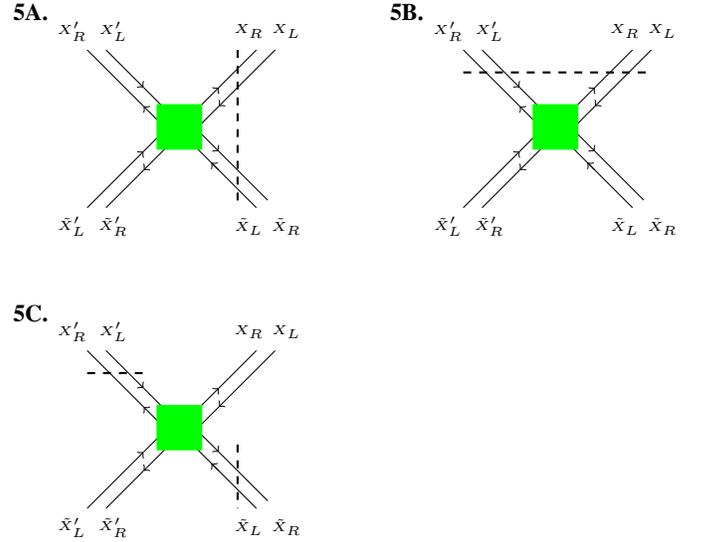

The conductances that we are interested in relate the current through the point $x \in [0, \ell]$ about the dashed lines shown in Fig. \ref{figfive} obtained in response to a constant electric field applied to a section of wire of length $\ell$.
The currents,
\begin{align}
\label{twotermcurrents}
J_{{\bf \hat{x}}}(x, \tau) & = {1 \over \sqrt{2} 2 \pi} \Big[ \Big({1 \over \sqrt{g}} + \sqrt{g}\Big) \Big(\partial_{z} \tilde{X}_L(z)- \bar{\partial}_{\bar{z}} X_R(\bar{z})\Big) \cr
& + \Big({1 \over \sqrt{g}} - \sqrt{g}\Big) \Big(\partial_{z} X_L(z) - \bar{\partial}_{\bar{z}} \tilde{X}_R(\bar{z})\Big)\Big], \cr
J_{{\bf \hat{y}}}(x, \tau) & = {1 \over \sqrt{2} 2 \pi} \Big[ \Big({1 \over \sqrt{g}} + \sqrt{g}\Big) \Big(\partial_{z}X'_L(z)- \bar{\partial}_{\bar{z}} X_R(\bar{z})\Big) \cr
& + \Big({1 \over \sqrt{g}} - \sqrt{g}\Big) \Big(\partial_{z} X_L(z) - \bar{\partial}_{\bar{z}} X'_R(\bar{z})\Big)\Big], \cr
J_{{\rm s}}(x, \tau) & = {1 \over \sqrt{2} 2 \pi} \Big[\Big({1 \over \sqrt{g}} + \sqrt{g}\Big) \Big(\partial_{z} \tilde{X}_L(z) + \partial_{z}X'_L(z)\Big) \cr
& - \Big({1 \over \sqrt{g}} - \sqrt{g}\Big) \Big(\bar{\partial}_{\bar{z}} \tilde{X}_R(\bar{z}) + \bar{\partial}_{\bar{z}} X'_R(\bar{z})\Big)\Big],
\end{align}
where $z = x + i \tau$.
Using the Kubo formula, we shall calculate the following three conductances at the fixed points previously described for $g>1$:
\begin{align}
\label{conductances}
G^{(A)}_{{\bf \hat{a}}, {\bf \hat{a}}} & = {e^2 \over \hbar \ell} \int_0^\ell dx' \lim_{\omega \rightarrow 0} \int_{- \infty}^\infty d \tau e^{i \omega \tau} {\langle J_{{\bf \hat{a}}}(x, \tau) J_{{\bf \hat{a}}}(x',0) \rangle_{A} \over \omega}, 
\end{align}
where $A = FC, PS$, or $FD$ denote the fully connected, partially-split, and fully disconnected fixed points and ${\bf \hat{a}} = {\bf \hat{x}}, {\bf \hat{y}}$, or ${\bf \hat{s}}$.

The first two conductances of Eqn. (\ref{twotermcurrents}) are sufficient to distinguish the fixed points and make particularly clear the symmetry present at the partially-split fixed point.
From the perspective of a fully connected geometry shown in Fig. \ref{figone}A, $G^{(FC)}_{{\bf \hat{x}}, {\bf \hat{x}}}$ measures the conductance along the bar, while $G^{(FC)}_{{\bf \hat{y}}, {\bf \hat{y}}}$ measures the conductance across the Hall bar.
$G^{(FC)}_{{\bf \hat{s}}, {\bf \hat{s}}}$ is a type of skew conductance where the potential on leads on diametrically-opposite sides of the point contact are raised relative to the other two.
Its value, however, is independent of the particular fixed point considered and we shall not consider it further.

The central results of this section are the following two-terminal electrical conductances:
\begin{align}
\label{conductancevalues}
G^{(FC)}_{{\bf \hat{x}}, {\bf \hat{x}}} & = G^{(FD)}_{{\bf \hat{y}}, {\bf \hat{y}}} = \Big({1 \over g} + g\Big) {e^2 \over h}, \cr
G^{(FD)}_{{\bf \hat{x}},{\bf \hat{x}}} & = G^{(FC)}_{{\bf \hat{y}}, {\bf \hat{y}}} = 0, \cr
G^{(PS)}_{{\bf \hat{x}}, {\bf \hat{x}}} & = G^{(PS)}_{{\bf \hat{y}}, {\bf \hat{y}}} = {1 \over g} {e^2 \over h}.
\end{align}
We note the equality of the conductances along and across the Hall bar at the partially-split fixed point.
Using either the mapping described in the previous section or an explicit calculation, the conductances for $g<1$ are obtained by substituting $g \leftrightarrow 1/g$ in Eqn. (\ref{conductancevalues}).
Interestingly, setting $e^2/h = 1$, the conductance at the partially-split fixed point is equal to the boundary scaling dimension of the operator ${\cal O}_{1,1}$ (${\cal O}'_{1,1}$) that drives the system towards the intermediate fixed point from either the fully connected or fully disconnected fixed points.

In order to calculate $G^{(A)}_{{\bf \hat{x}}, {\bf \hat{x}}}$, for instance, we need the following correlation functions.
Note that only a subset of the non-zero correlation functions are listed with the additional correlators obtained by methods exactly similar to those that we state below.
Correlation functions containing purely (anti-)holomorphic fields are not generally affected by the boundary conditions at the point contact:
\begin{align}
\label{currentcorrelators}
\langle \bar{\partial}_{\bar{z}} X_R(\bar{z}) \bar{\partial}_{\bar{z}'} X_R(\bar{z}') \rangle_A & = {1 \over (\bar{z} - \bar{z}')^2}, \cr
\langle \partial_{z} X_L(z) \partial_{z'} X_L(z') \rangle_A & = {1 \over (z - z')^2},
\end{align}
and similarly for other two-point correlators of pairs of purely (anti-)holomorphic fields.
Correlation functions between two different (anti-)holomorphic fields vanish.
However, correlators between a holomorphic field and an anti-holomorphic field {\it do} depend upon the specific boundary conditions at the point contact:
\begin{align}
\label{holantiholcorr}
\langle \bar{\partial}_{\bar{z}} X_R(\bar{z}) \partial_{z'} X'_L(z') \rangle_A & = {1 \over (z' + \bar{z})^2} \Big(\delta_{A, FC} + {1 \over 2}\delta_{A, PS}\Big), \cr
\langle \bar{\partial}_{\bar{z}} X_R(\bar{z}) \partial_{z'} X_L(z') \rangle_A & = - {1 \over 2(z' + \bar{z})^2} \delta_{A, PS}, \cr
\langle \bar{\partial}_{\bar{z}} X_R(\bar{z}) \partial_{z'} \tilde{X}_L(z') \rangle_A & = {1 \over (z' + \bar{z})^2} \Big(\delta_{A, FD} + {1 \over 2}\delta_{A, PS}\Big), \cr
\langle \bar{\partial}_{\bar{z}} X_R(\bar{z}) \partial_{z'} \tilde{X}'_L(z') \rangle_A & = {1 \over 2(z' + \bar{z})^2} \delta_{A, PS}.
\end{align}
In deriving Eqn. (\ref{holantiholcorr}), we used the field redefinition in Eqn. (\ref{leftrightdef}) and the fact that:
\begin{align}
\langle \bar{\partial}_{\bar{z}} R_i (\bar{z}) \partial_{z'} L_j(z') \rangle_A = \pm \delta_{i j} {1 \over (z' + \bar{z})^2},
\end{align}
if $R_i (0, \tau) = \pm L_j(0,\tau)$ at the point contact.

We are now ready to verify the conductances quoted in Eqn. (\ref{conductancevalues}).
The calculations of $G^{(A)}_{{\bf \hat{x}}, {\bf \hat{x}}}$ and $G^{(A)}_{{\bf \hat{y}}, {\bf \hat{y}}}$ are related by the replacements, $\tilde{X}_{R,L} \leftrightarrow X'_{R,L}$ and $\delta_{A, FD} \leftrightarrow \delta_{A, FC}$.
(The calculation of $G^{(A)}_{{\bf \hat{s}}, {\bf \hat{s}}}$ proceeds analogously.)
Therefore, we only show the calculation of $G^{(A)}_{{\bf \hat{x}}, {\bf \hat{x}}}$ below.

The correlation function in the top line of Eqn. (\ref{conductances}) factors into twelve non-zero terms:
\begin{widetext}
\begin{align}
\label{details}
\langle J_{{\bf \hat{x}}}(x, \tau) J_{{\bf \hat{x}}}(x',0) \rangle_{A} & = {1 \over 8 \pi^2} \Big[({1 \over \sqrt{g}} + \sqrt{g})^2 \Big(\langle \bar{\partial}_{\bar{z}} X_R(\bar{z}) \bar{\partial}_{\bar{z}'} X_R(\bar{z}') \rangle_A - \langle \bar{\partial}_{\bar{z}} X_R(\bar{z}) \partial_{z'} \tilde{X}_L(z') \rangle_A - \langle \partial_{z} \tilde{X}_L(z) \bar{\partial}_{\bar{z}'} X_R(\bar{z}') \rangle_A \cr
& + \langle \partial_{z} \tilde{X}_L(z) \partial_{z'} \tilde{X}_L(z') \rangle_A\Big) + ({1 \over \sqrt{g}} - \sqrt{g})^2 \Big(\langle \bar{\partial}_{\bar{z}} \tilde{X}_R(\bar{z}) \bar{\partial}_{\bar{z}'} \tilde{X}_R(\bar{z}') \rangle_A - \langle \bar{\partial}_{\bar{z}} \tilde{X}_R(\bar{z}) \partial_{z'} X_L(z') \rangle_A \cr
& - \langle \partial_{z} X_L(z) \bar{\partial}_{\bar{z}'} \tilde{X}_R(\bar{z}') \rangle_A + \langle \partial_{z} X_L(z) \partial_{z'} X_L(z') \rangle_A\Big) - ({1 \over g} - g)\Big(\langle \bar{\partial}_{\bar{z}} X_R(\bar{z}) \partial_{z'} X_L(z') \rangle_A \cr
& + \langle \partial_{z} X_L(z) \bar{\partial}_{\bar{z}'} X_R(\bar{z}') \rangle_A + \langle \bar{\partial}_{\bar{z}} \tilde{X}_R(\bar{z}) \partial_{z'} \tilde{X}_L(z') \rangle_A + \langle \partial_{z} \tilde{X}_L(z) \bar{\partial}_{\bar{z}'} \tilde{X}_R(\bar{z}') \rangle_A \Big],
\end{align}
\end{widetext}
where $z = x + i \tau$ and $z' = x'$.
Making use of Eqn. (\ref{currentcorrelators}), (\ref{holantiholcorr}) in the above equation, and substituting into the Kubo formula, Eqn. (\ref{conductances}), we encounter integrals of the form:
\begin{align}
\label{fouriertrans}
\int_{- \infty}^\infty d \tau {e^{i \omega \tau} \over \Big(\pm i \tau + (x - x')\Big)^2} = 2 \pi \omega e^{- |x-x'| \omega} \Theta\Big(\pm(x-x')\Big),
\end{align}
where $\Theta(x) = 1$ for $x >0$ and $\Theta(x) = 0$ for $x<0$ is the Heaviside step function.
(The non-zero difference $|x-x'|$ functions as an UV regulator of the $\tau$-integral.)
Taking the DC $\omega \rightarrow 0$ limit and integrating over $x'$, we find:
\begin{align}
G^{(A)}_{{\bf \hat{x}}, {\bf \hat{x}}} = {e^2 \over h} \Big( ({1 \over g} + g) - ({1 \over g} + g) \delta_{A, FD} - g \delta_{A, PS}\Big).
\end{align}
Thus, we have checked the value of $G^{(A)}_{{\bf \hat{x}}, {\bf \hat{x}}}$ quoted in Eqn. (\ref{conductancevalues}).
The calculation of $G^{(A)}_{{\bf \hat{y}}, {\bf \hat{y}}}$ proceeds in precisely the same way with the substitution $\tilde{X}_{R,L} \leftrightarrow X'_{L,R}$ in Eqn. (\ref{details}).




\section{Summary}
\label{summary}

In this note, we have studied how the bosonic integer quantum Hall state responds to two distinct perturbations: disorder and point-contact tunneling.
The bosonic IQH state is stable as long as charge conservation is maintained, however, the two-terminal electrical Hall conductance inferred from the conductance of its edge modes leads to a value that depends continuously on a particular edge-mode interaction parameter that we denoted by $g$.
When inter-mode equilibration occurs via impurities, the interaction parameterized by $g$ renormalizes towards the value $g=1$ at which the two-terminal electrical conductance $G$ takes the universal value of $2 e^2/h$ and the leading point-contact tunneling perturbation is marginal.

Equilibration via interactions induced by impurities along the edge is only a relevant perturbation when $g$ is within a certain range of values.
Outside of the domain of attraction of the strong-disorder fixed point, these impurity-mediated interactions are irrelevant and so equilibration does not occur.
The two-terminal electrical conductance then depends upon the edge interaction parameter $g$.

When the impurity-mediated interactions are irrelevant or when the temperature cuts off the RG flow towards the strong-disorder fixed point, we may consider how a point contact affects the bosonic IQH state.
Away from the $g=1$ point, a point contact drives a boundary RG trajectory to a single, isolated IR stable fixed point where the Hall bar partially splits into two pieces.
The partially-split fixed point is characterized by equal two-terminal electrical conductances both along and across the Hall bar.

This point-contact induced partial splitting should be contrasted with other short-range entangled states whose RG trajectories are drawn in Fig. \ref{figtwo}.
Fermionic IQH states do not split under renormalization group flow as a result of perturbation by a point contact -- this perturbation is exactly marginal -- and instead are described by a line of fixed points.
All point contact perturbations of the $E_8$ state of bosons are strictly irrelevant and do not affect the low energy physics at weak coupling.\cite{Kitaevanyonsexact, Kitaev11, LuVishwanathcsclass, PlamadealaE8}

The time-reversal invariant spin Hall insulator shows behavior similar to its bosonic counterpart studied in this paper.
In contrast to the spin Hall insulator, the bosonic IQH state generically admits a relevant point contact perturbation.

Long-range entangled states generally {\it do} split as a result of a relevant point-contact perturbation.
Thus, the bosonic integer quantum Hall state in the clean limit and the spin Hall insulator fit somewhere in between fermionic IQH states and long-range entangled states in terms of their response to a point contact.

This behavior is in keeping with the general interpretation of point-contact induced RG flows as dynamical loss of entropy.\cite{FFNholographic}
Because ``there is no entropy to lose," short-range entangled states cannot RG flow between fully connected and fully disconnected fixed points; there must either be no flow, or a flow to or from an intermediate fixed point where the Hall bar has partially split.
The RG diagrams in Fig. \ref{figtwo} (ii) and (iii) of the bosonic integer quantum Hall effect reflect this behavior.

Boundary conformal fixed points have a beautiful description in terms of boundary states.\cite{Cardybcfusionverlinde}
Indeed, the Affleck-Ludwig boundary entropy\cite{AffleckLudwigentropy} characterizing the boundary renormalization group flow naturally emerges within this formalism.
It is an interesting open question to consider how the boundary state formalism could shed light on the properties of the IR unstable boundary fixed point obtained from the point-contact perturbed $E_8$ state.

In the clean limit, the edge interaction parameter $g$ was taken to be constant and equal on both edges which interacted via the point contact.
It might be interesting to consider a more general, but related, setup described by a junction of four quantum wires each with an independent interaction parameters $g_i$.
(See Ref. [\onlinecite{PhysRevB.86.075451}] for the cases of two- and three-wire junctions.)

The one-to-many nature of the association between the bulk topological order of an abelian Hall state and its edge modes has been emphasized recently:\cite{PlamadealaE8, generalstableequivalence} the same bulk state can admit more than one distinct edge phase.
Ignoring symmetry, the spin-Hall insulator admits an edge transition to a phase in which the low energy edge excitations are identical to those of the bosonic integer quantum Hall state.
While this transition necessarily breaks time-reversal, it is entirely possible for it to be the most relevant instability of the spin-Hall edge modes (when time-reversal is allowed to be broken) by fine-tuning edge interaction parameters (the analog of what we called $V_{IJ}$).
It would be interesting to consider in further work the relation between these two edge phases.

\acknowledgments

We are grateful to Jennifer Cano, Chetan Nayak, and Eugeniu Plamadeala for helpful discussions.
This research was supported in part by the National Science Foundation under
Grant DMR-1101912 (M.P.A.F.) and by the Caltech Institute of Quantum Information
and Matter, an NSF Physics Frontiers Center with support of
the Gordon and Betty Moore Foundation (M.P.A.F.).

\bibliography{BIQHtunneling}

\begin{thebibliography}{41}
\expandafter\ifx\csname natexlab\endcsname\relax\def\natexlab#1{#1}\fi
\expandafter\ifx\csname bibnamefont\endcsname\relax
  \def\bibnamefont#1{#1}\fi
\expandafter\ifx\csname bibfnamefont\endcsname\relax
  \def\bibfnamefont#1{#1}\fi
\expandafter\ifx\csname citenamefont\endcsname\relax
  \def\citenamefont#1{#1}\fi
\expandafter\ifx\csname url\endcsname\relax
  \def\url#1{\texttt{#1}}\fi
\expandafter\ifx\csname urlprefix\endcsname\relax\def\urlprefix{URL }\fi
\providecommand{\bibinfo}[2]{#2}
\providecommand{\eprint}[2][]{\url{#2}}

\bibitem[{\citenamefont{Kane and Fisher}(1992)}]{kanefishertunnel}
\bibinfo{author}{\bibfnamefont{C.~L.} \bibnamefont{Kane}} \bibnamefont{and}
  \bibinfo{author}{\bibfnamefont{M.~P.~A.} \bibnamefont{Fisher}},
  \bibinfo{journal}{Phys. Rev. B} \textbf{\bibinfo{volume}{46}},
  \bibinfo{pages}{15233} (\bibinfo{year}{1992}).

\bibitem[{\citenamefont{Fendley
  et~al.}(1995{\natexlab{a}})\citenamefont{Fendley, Ludwig, and
  Saleur}}]{FLSshort}
\bibinfo{author}{\bibfnamefont{P.}~\bibnamefont{Fendley}},
  \bibinfo{author}{\bibfnamefont{A.~W.~W.} \bibnamefont{Ludwig}},
  \bibnamefont{and} \bibinfo{author}{\bibfnamefont{H.}~\bibnamefont{Saleur}},
  \bibinfo{journal}{Phys. Rev. Lett.} \textbf{\bibinfo{volume}{74}},
  \bibinfo{pages}{3005} (\bibinfo{year}{1995}{\natexlab{a}}).

\bibitem[{\citenamefont{Fendley
  et~al.}(1995{\natexlab{b}})\citenamefont{Fendley, Ludwig, and
  Saleur}}]{FLSlong}
\bibinfo{author}{\bibfnamefont{P.}~\bibnamefont{Fendley}},
  \bibinfo{author}{\bibfnamefont{A.~W.~W.} \bibnamefont{Ludwig}},
  \bibnamefont{and} \bibinfo{author}{\bibfnamefont{H.}~\bibnamefont{Saleur}},
  \bibinfo{journal}{Phys. Rev. B} \textbf{\bibinfo{volume}{52}},
  \bibinfo{pages}{8934} (\bibinfo{year}{1995}{\natexlab{b}}).

\bibitem[{\citenamefont{Wong and Affleck}(1994)}]{wongafflecktunnel}
\bibinfo{author}{\bibfnamefont{E.}~\bibnamefont{Wong}} \bibnamefont{and}
  \bibinfo{author}{\bibfnamefont{I.}~\bibnamefont{Affleck}},
  \bibinfo{journal}{Nucl.Phys. B} \textbf{\bibinfo{volume}{417}},
  \bibinfo{pages}{403} (\bibinfo{year}{1994}).

\bibitem[{\citenamefont{Cardy}(1989)}]{Cardybcfusionverlinde}
\bibinfo{author}{\bibfnamefont{J.~L.} \bibnamefont{Cardy}},
  \bibinfo{journal}{Nucl.Phys. B} \textbf{\bibinfo{volume}{324}},
  \bibinfo{pages}{581} (\bibinfo{year}{1989}).

\bibitem[{\citenamefont{Affleck and Ludwig}(1991)}]{AffleckLudwigentropy}
\bibinfo{author}{\bibfnamefont{I.}~\bibnamefont{Affleck}} \bibnamefont{and}
  \bibinfo{author}{\bibfnamefont{A.~W.} \bibnamefont{Ludwig}},
  \bibinfo{journal}{Phys.Rev.Lett.} \textbf{\bibinfo{volume}{67}},
  \bibinfo{pages}{161} (\bibinfo{year}{1991}).

\bibitem[{\citenamefont{Zamolodchikov}(1986)}]{Zamolodchikovctheorem}
\bibinfo{author}{\bibfnamefont{A.}~\bibnamefont{Zamolodchikov}},
  \bibinfo{journal}{JETP Lett.} \textbf{\bibinfo{volume}{43}},
  \bibinfo{pages}{730} (\bibinfo{year}{1986}).

\bibitem[{\citenamefont{Friedan and Konechny}(2004)}]{FriedanKonechny}
\bibinfo{author}{\bibfnamefont{D.}~\bibnamefont{Friedan}} \bibnamefont{and}
  \bibinfo{author}{\bibfnamefont{A.}~\bibnamefont{Konechny}},
  \bibinfo{journal}{Phys.Rev.Lett.} \textbf{\bibinfo{volume}{93}},
  \bibinfo{pages}{030402} (\bibinfo{year}{2004}).

\bibitem[{\citenamefont{Fendley et~al.}(2007)\citenamefont{Fendley, Fisher, and
  Nayak}}]{FFNholographic}
\bibinfo{author}{\bibfnamefont{P.}~\bibnamefont{Fendley}},
  \bibinfo{author}{\bibfnamefont{M.~P.} \bibnamefont{Fisher}},
  \bibnamefont{and} \bibinfo{author}{\bibfnamefont{C.}~\bibnamefont{Nayak}},
  \bibinfo{journal}{J.Statist.Phys.} \textbf{\bibinfo{volume}{126}},
  \bibinfo{pages}{1111} (\bibinfo{year}{2007}).

\bibitem[{\citenamefont{Hamma et~al.}(2005)\citenamefont{Hamma, Ionicioiu, and
  Zanardi}}]{hammakitaev}
\bibinfo{author}{\bibfnamefont{A.}~\bibnamefont{Hamma}},
  \bibinfo{author}{\bibfnamefont{R.}~\bibnamefont{Ionicioiu}},
  \bibnamefont{and} \bibinfo{author}{\bibfnamefont{P.}~\bibnamefont{Zanardi}},
  \bibinfo{journal}{Physics Letters A} \textbf{\bibinfo{volume}{337}},
  \bibinfo{pages}{22} (\bibinfo{year}{2005}).

\bibitem[{\citenamefont{Kitaev and Preskill}(2006)}]{KitaevPreskillentropy}
\bibinfo{author}{\bibfnamefont{A.}~\bibnamefont{Kitaev}} \bibnamefont{and}
  \bibinfo{author}{\bibfnamefont{J.}~\bibnamefont{Preskill}},
  \bibinfo{journal}{Phys.Rev.Lett.} \textbf{\bibinfo{volume}{96}},
  \bibinfo{pages}{110404} (\bibinfo{year}{2006}).

\bibitem[{\citenamefont{Levin and Wen}(2006)}]{LevinWenentropy}
\bibinfo{author}{\bibfnamefont{M.}~\bibnamefont{Levin}} \bibnamefont{and}
  \bibinfo{author}{\bibfnamefont{X.-G.} \bibnamefont{Wen}},
  \bibinfo{journal}{Phys.Rev.Lett.} \textbf{\bibinfo{volume}{96}},
  \bibinfo{pages}{110405} (\bibinfo{year}{2006}).

\bibitem[{\citenamefont{Kane and Fisher}(1997)}]{kanefisherthermal}
\bibinfo{author}{\bibfnamefont{C.~L.} \bibnamefont{Kane}} \bibnamefont{and}
  \bibinfo{author}{\bibfnamefont{M.~P.~A.} \bibnamefont{Fisher}},
  \bibinfo{journal}{Phys. Rev. B} \textbf{\bibinfo{volume}{55}},
  \bibinfo{pages}{15832} (\bibinfo{year}{1997}).

\bibitem[{\citenamefont{Kitaev}(2006)}]{Kitaevanyonsexact}
\bibinfo{author}{\bibfnamefont{A.}~\bibnamefont{Kitaev}},
  \bibinfo{journal}{Annals Phys.} \textbf{\bibinfo{volume}{321}},
  \bibinfo{pages}{2} (\bibinfo{year}{2006}).

\bibitem[{Kit()}]{Kitaev11}
\bibinfo{note}{A. Kitaev, http://online.kitp.ucsb.edu/online/topomat11/kitaev.}

\bibitem[{\citenamefont{Lu and Vishwanath}(2012)}]{LuVishwanathcsclass}
\bibinfo{author}{\bibfnamefont{Y.-M.} \bibnamefont{Lu}} \bibnamefont{and}
  \bibinfo{author}{\bibfnamefont{A.}~\bibnamefont{Vishwanath}},
  \bibinfo{journal}{Phys.Rev. B} \textbf{\bibinfo{volume}{86}},
  \bibinfo{pages}{125119} (\bibinfo{year}{2012}).

\bibitem[{\citenamefont{Plamadeala et~al.}(2013)\citenamefont{Plamadeala,
  Mulligan, and Nayak}}]{PlamadealaE8}
\bibinfo{author}{\bibfnamefont{E.}~\bibnamefont{Plamadeala}},
  \bibinfo{author}{\bibfnamefont{M.}~\bibnamefont{Mulligan}}, \bibnamefont{and}
  \bibinfo{author}{\bibfnamefont{C.}~\bibnamefont{Nayak}},
  \bibinfo{journal}{Phys. Rev. B} \textbf{\bibinfo{volume}{88}},
  \bibinfo{pages}{045131} (\bibinfo{year}{2013}).

\bibitem[{\citenamefont{Fu and Kane}(2007)}]{FuKaneinversion}
\bibinfo{author}{\bibfnamefont{L.}~\bibnamefont{Fu}} \bibnamefont{and}
  \bibinfo{author}{\bibfnamefont{C.~L.} \bibnamefont{Kane}},
  \bibinfo{journal}{Phys. Rev. B} \textbf{\bibinfo{volume}{76}},
  \bibinfo{pages}{045302} (\bibinfo{year}{2007}).

\bibitem[{\citenamefont{Qi et~al.}(2008)\citenamefont{Qi, Hughes, and
  Zhang}}]{QHZtop}
\bibinfo{author}{\bibfnamefont{X.-L.} \bibnamefont{Qi}},
  \bibinfo{author}{\bibfnamefont{T.~L.} \bibnamefont{Hughes}},
  \bibnamefont{and} \bibinfo{author}{\bibfnamefont{S.-C.} \bibnamefont{Zhang}},
  \bibinfo{journal}{Phys. Rev. B} \textbf{\bibinfo{volume}{78}},
  \bibinfo{pages}{195424} (\bibinfo{year}{2008}).

\bibitem[{\citenamefont{Chen et~al.}(2011)\citenamefont{Chen, Liu, and
  Wen}}]{xieliuwen}
\bibinfo{author}{\bibfnamefont{X.}~\bibnamefont{Chen}},
  \bibinfo{author}{\bibfnamefont{Z.-X.} \bibnamefont{Liu}}, \bibnamefont{and}
  \bibinfo{author}{\bibfnamefont{X.-G.} \bibnamefont{Wen}},
  \bibinfo{journal}{Phys. Rev. B} \textbf{\bibinfo{volume}{84}},
  \bibinfo{pages}{235141} (\bibinfo{year}{2011}).

\bibitem[{\citenamefont{{Chen} et~al.}()\citenamefont{{Chen}, {Gu}, {Liu}, and
  {Wen}}}]{Chen11b}
\bibinfo{author}{\bibfnamefont{X.}~\bibnamefont{{Chen}}},
  \bibinfo{author}{\bibfnamefont{Z.-C.} \bibnamefont{{Gu}}},
  \bibinfo{author}{\bibfnamefont{Z.-X.} \bibnamefont{{Liu}}}, \bibnamefont{and}
  \bibinfo{author}{\bibfnamefont{X.-G.} \bibnamefont{{Wen}}},
  \bibinfo{note}{arXiv:1106.4772}.

\bibitem[{\citenamefont{Hou et~al.}(2009)\citenamefont{Hou, Kim, and
  Chamon}}]{houkimchamon}
\bibinfo{author}{\bibfnamefont{C.-Y.} \bibnamefont{Hou}},
  \bibinfo{author}{\bibfnamefont{E.-A.} \bibnamefont{Kim}}, \bibnamefont{and}
  \bibinfo{author}{\bibfnamefont{C.}~\bibnamefont{Chamon}},
  \bibinfo{journal}{Phys. Rev. Lett.} \textbf{\bibinfo{volume}{102}},
  \bibinfo{pages}{076602} (\bibinfo{year}{2009}).

\bibitem[{\citenamefont{Teo and Kane}(2009)}]{teokanepointcontact}
\bibinfo{author}{\bibfnamefont{J.~C.~Y.} \bibnamefont{Teo}} \bibnamefont{and}
  \bibinfo{author}{\bibfnamefont{C.~L.} \bibnamefont{Kane}},
  \bibinfo{journal}{Phys. Rev. B} \textbf{\bibinfo{volume}{79}},
  \bibinfo{pages}{235321} (\bibinfo{year}{2009}).

\bibitem[{\citenamefont{Senthil and Levin}(2013)}]{LevinSenthilbosons}
\bibinfo{author}{\bibfnamefont{T.}~\bibnamefont{Senthil}} \bibnamefont{and}
  \bibinfo{author}{\bibfnamefont{M.}~\bibnamefont{Levin}},
  \bibinfo{journal}{Phys. Rev. Lett.} \textbf{\bibinfo{volume}{110}},
  \bibinfo{pages}{046801} (\bibinfo{year}{2013}).

\bibitem[{\citenamefont{Ye and Wen}(2013)}]{YeWen}
\bibinfo{author}{\bibfnamefont{P.}~\bibnamefont{Ye}} \bibnamefont{and}
  \bibinfo{author}{\bibfnamefont{X.-G.} \bibnamefont{Wen}},
  \bibinfo{journal}{Phys. Rev. B} \textbf{\bibinfo{volume}{87}},
  \bibinfo{pages}{195128} (\bibinfo{year}{2013}).

\bibitem[{\citenamefont{Kane et~al.}(1994)\citenamefont{Kane, Fisher, and
  Polchinski}}]{KFPrandom}
\bibinfo{author}{\bibfnamefont{C.~L.} \bibnamefont{Kane}},
  \bibinfo{author}{\bibfnamefont{M.~P.~A.} \bibnamefont{Fisher}},
  \bibnamefont{and}
  \bibinfo{author}{\bibfnamefont{J.}~\bibnamefont{Polchinski}},
  \bibinfo{journal}{Phys. Rev. Lett.} \textbf{\bibinfo{volume}{72}},
  \bibinfo{pages}{4129} (\bibinfo{year}{1994}).

\bibitem[{\citenamefont{Kane and Fisher}(1995)}]{KFrandom}
\bibinfo{author}{\bibfnamefont{C.~L.} \bibnamefont{Kane}} \bibnamefont{and}
  \bibinfo{author}{\bibfnamefont{M.~P.~A.} \bibnamefont{Fisher}},
  \bibinfo{journal}{Phys. Rev. B} \textbf{\bibinfo{volume}{51}},
  \bibinfo{pages}{13449} (\bibinfo{year}{1995}).

\bibitem[{\citenamefont{Wen}(1995)}]{wenreview}
\bibinfo{author}{\bibfnamefont{X.~G.} \bibnamefont{Wen}},
  \bibinfo{journal}{Adv. Phys.} \textbf{\bibinfo{volume}{44}},
  \bibinfo{pages}{405} (\bibinfo{year}{1995}).

\bibitem[{\citenamefont{Elitzur et~al.}(1989)\citenamefont{Elitzur, Moore,
  Schwimmer, and Seiberg}}]{Elitzur89}
\bibinfo{author}{\bibfnamefont{S.}~\bibnamefont{Elitzur}},
  \bibinfo{author}{\bibfnamefont{G.~W.} \bibnamefont{Moore}},
  \bibinfo{author}{\bibfnamefont{A.}~\bibnamefont{Schwimmer}},
  \bibnamefont{and} \bibinfo{author}{\bibfnamefont{N.}~\bibnamefont{Seiberg}},
  \bibinfo{journal}{Nucl. Phys. B} \textbf{\bibinfo{volume}{326}},
  \bibinfo{pages}{108} (\bibinfo{year}{1989}).

\bibitem[{\citenamefont{Wen}(1992)}]{Wen92edgereview}
\bibinfo{author}{\bibfnamefont{X.-G.} \bibnamefont{Wen}},
  \bibinfo{journal}{Int. J. Mod. Phys. B} \textbf{\bibinfo{volume}{6}},
  \bibinfo{pages}{1711} (\bibinfo{year}{1992}).

\bibitem[{\citenamefont{Haldane}(1995)}]{Haldanestability}
\bibinfo{author}{\bibfnamefont{F.~D.~M.} \bibnamefont{Haldane}},
  \bibinfo{journal}{Phys. Rev. Lett.} \textbf{\bibinfo{volume}{74}},
  \bibinfo{pages}{2090} (\bibinfo{year}{1995}).

\bibitem[{\citenamefont{Levin}(2013)}]{Levin13}
\bibinfo{author}{\bibfnamefont{M.}~\bibnamefont{Levin}},
  \bibinfo{journal}{Phys. Rev. X} \textbf{\bibinfo{volume}{3}},
  \bibinfo{pages}{021009} (\bibinfo{year}{2013}).

\bibitem[{wan()}]{wangwen12}
\bibinfo{note}{Juven Wang and Xiao-Gang Wen, arXiv:1212.4863}.

\bibitem[{\citenamefont{Barkeshli et~al.}(2013)\citenamefont{Barkeshli, Jian,
  and Qi}}]{barkeshlijianqiboundary}
\bibinfo{author}{\bibfnamefont{M.}~\bibnamefont{Barkeshli}},
  \bibinfo{author}{\bibfnamefont{C.-M.} \bibnamefont{Jian}}, \bibnamefont{and}
  \bibinfo{author}{\bibfnamefont{X.-L.} \bibnamefont{Qi}},
  \bibinfo{journal}{Phys. Rev. B} \textbf{\bibinfo{volume}{88}},
  \bibinfo{pages}{235103} (\bibinfo{year}{2013}).

\bibitem[{\citenamefont{Maslov and Stone}(1995)}]{maslovstone}
\bibinfo{author}{\bibfnamefont{D.~L.} \bibnamefont{Maslov}} \bibnamefont{and}
  \bibinfo{author}{\bibfnamefont{M.}~\bibnamefont{Stone}},
  \bibinfo{journal}{Phys. Rev. B} \textbf{\bibinfo{volume}{52}},
  \bibinfo{pages}{R5539} (\bibinfo{year}{1995}).

\bibitem[{\citenamefont{Safi and Schulz}(1995)}]{SafiSchulz}
\bibinfo{author}{\bibfnamefont{I.}~\bibnamefont{Safi}} \bibnamefont{and}
  \bibinfo{author}{\bibfnamefont{H.~J.} \bibnamefont{Schulz}},
  \bibinfo{journal}{Phys. Rev. B} \textbf{\bibinfo{volume}{52}},
  \bibinfo{pages}{R17040} (\bibinfo{year}{1995}).

\bibitem[{\citenamefont{Landauer}(1970)}]{Landauer}
\bibinfo{author}{\bibfnamefont{R.}~\bibnamefont{Landauer}},
  \bibinfo{journal}{Philosophical Magazine} \textbf{\bibinfo{volume}{21}},
  \bibinfo{pages}{863} (\bibinfo{year}{1970}).

\bibitem[{\citenamefont{B\"uttiker}(1986)}]{Buttiker}
\bibinfo{author}{\bibfnamefont{M.}~\bibnamefont{B\"uttiker}},
  \bibinfo{journal}{Phys. Rev. Lett.} \textbf{\bibinfo{volume}{57}},
  \bibinfo{pages}{1761} (\bibinfo{year}{1986}).

\bibitem[{\citenamefont{Giamarchi and Schulz}(1988)}]{GiamarchiSchulz}
\bibinfo{author}{\bibfnamefont{T.}~\bibnamefont{Giamarchi}} \bibnamefont{and}
  \bibinfo{author}{\bibfnamefont{H.~J.} \bibnamefont{Schulz}},
  \bibinfo{journal}{Phys. Rev. B} \textbf{\bibinfo{volume}{37}},
  \bibinfo{pages}{325} (\bibinfo{year}{1988}).

\bibitem[{\citenamefont{Hou et~al.}(2012)\citenamefont{Hou, Rahmani, Feiguin,
  and Chamon}}]{PhysRevB.86.075451}
\bibinfo{author}{\bibfnamefont{C.-Y.} \bibnamefont{Hou}},
  \bibinfo{author}{\bibfnamefont{A.}~\bibnamefont{Rahmani}},
  \bibinfo{author}{\bibfnamefont{A.~E.} \bibnamefont{Feiguin}},
  \bibnamefont{and} \bibinfo{author}{\bibfnamefont{C.}~\bibnamefont{Chamon}},
  \bibinfo{journal}{Phys. Rev. B} \textbf{\bibinfo{volume}{86}},
  \bibinfo{pages}{075451} (\bibinfo{year}{2012}).

\bibitem[{\citenamefont{Cano et~al.}(2014)\citenamefont{Cano, Cheng, Mulligan,
  Nayak, Plamadeala, and Yard}}]{generalstableequivalence}
\bibinfo{author}{\bibfnamefont{J.}~\bibnamefont{Cano}},
  \bibinfo{author}{\bibfnamefont{M.}~\bibnamefont{Cheng}},
  \bibinfo{author}{\bibfnamefont{M.}~\bibnamefont{Mulligan}},
  \bibinfo{author}{\bibfnamefont{C.}~\bibnamefont{Nayak}},
  \bibinfo{author}{\bibfnamefont{E.}~\bibnamefont{Plamadeala}},
  \bibnamefont{and} \bibinfo{author}{\bibfnamefont{J.}~\bibnamefont{Yard}},
  \bibinfo{journal}{Phys. Rev. B} \textbf{\bibinfo{volume}{89}},
  \bibinfo{pages}{115116} (\bibinfo{year}{2014}).

\end{thebibliography}

\end{document}